\documentclass[a4paper,11pt]{article} 
\usepackage{jcappub} 
\pdfoutput=1 
\usepackage[utf8]{inputenc} 
\usepackage{subcaption} 
\usepackage{multirow}
\usepackage{threeparttable} 
\usepackage[table]{xcolor} 
\usepackage[noabbrev]{cleveref} 
\Crefname{equation}{Eq.}{Eqs.} 
\usepackage{aasmacros} 
\usepackage{xcolor} 
\usepackage[normalem]{ulem} 

\title{\boldmath{Primordial power spectrum reconstructions from BOSS + eBOSS}}

\author[1,2]{Guillermo Martínez-Somonte,}
\author[3,4,5]{Héctor Gil-Marín,}
\author[1]{Airam Marcos-Caballero,}
\author[1]{Enrique Martínez-González}

\affiliation[1]{Instituto de Física de Cantabria, CSIC-Universidad de Cantabria,\\Avenida de los Castros s/n, E-39005 Santander, Spain}
\affiliation[2]{Departamento de Física Moderna, Universidad de Cantabria,\\Avenida de los Castros s/n, E-39005 Santander, Spain}
\affiliation[3]{Departament de Física Quàntica i Astrofísica, Universitat de Barcelona,\
Martí i Franquès 1, E-08028 Barcelona, Spain}
\affiliation[4]{Institut d’Estudis Espacials de Catalunya (IEEC),\
08034 Barcelona, Spain}
\affiliation[5]{Institut de Ciències del Cosmos (ICCUB), Universitat de Barcelona (UB),\
c. Martí i Franquès 1, 08028 Barcelona, Spain}

\emailAdd{gmsomonte@ifca.es}
\emailAdd{hectorgil@icc.ub.edu}
\emailAdd{marcos@ifca.es}
\emailAdd{martinez@ifca.es}

\abstract{We reconstruct the primordial power spectrum $P_{\mathcal{R}}(k)$ from the BOSS DR 12 LRG and eBOSS DR 16 QSO catalogs with a non-parametric Bayesian method. The $P_{\mathcal{R}}(k)$ is reconstructed by linearly interpolating $N$ knots in the $\{ \log k, \log P_{\mathcal{R}}(k) \}$ plane. We use a parametric model to describe the galaxy power spectra of the BOSS+eBOSS catalogs, assuming any power-law deviations and BAO contributions separately from the matter power spectrum template, composed of seven parameters $\Theta_{\text{model}}$. This parametric model enables us to reconstruct $P_{\mathcal{R}}(k)$ at non-linear scales, reaching $k = 0.3 \text{ h} \text{ Mpc}^{-1}$. The method is validated by applying it to different Primordial Features (PF) templates and by recovering the input power law of \textsc{MD-Patchy} and \textsc{EZmock} mock catalogs, representative of the BOSS and eBOSS data. These mocks provide additional information on $\Theta_{\text{model}}$, enabling us to impose Gaussian correlated priors on $\Theta_{\text{model}}$. This prior set allows us to reconstruct $P_{\mathcal{R}}(k)$ more precisely and to alleviate the degeneracies between the model and knot parameters. The results for both individual and combined $z$-bins and galactic caps of the BOSS and eBOSS catalogs are consistent, showing no evidence of the presence of PF in $P_{\mathcal{R}}(k)$ and pointing to a quasi-scale-invariant power law as the preferred model for $P_{\mathcal{R}}(k)$, as predicted by most slow-roll inflationary models. With a different prior set that relaxes the Gaussian constraints on $\Theta_{\text{model}}$ and imposes Planck-based priors on the extreme knots, the results also favor the power law. From the knot reconstructions, we robustly constrain the spectral index $n_s = 0.976 \pm 0.021$, compatible with the Planck value.}


\begin{document}

\maketitle
\flushbottom 

\section{Introduction}
\label{sec:Introduction}

Primordial cosmological perturbations set the initial conditions for the formation of the Large-Scale Structure (LSS) and are commonly characterized by the primordial curvature power spectrum, $P_{\mathcal{R}}(k)$, whose shape encodes the physics of the early Universe, particularly during inflation. Within the standard $\Lambda$CDM paradigm, these perturbations are adiabatic and have statistical properties that are very close to Gaussian, while $P_{\mathcal{R}}(k)$ is well described by a nearly scale-invariant power-law, as generically expected in slow-roll inflation and confirmed by Cosmic Microwave Background (CMB) observations \cite{Planck18Parameters}. It is therefore usually described by two parameters, the amplitude of perturbations at a certain pivot scale, $A_s$, and the spectral index $n_s$ of the primordial comoving curvature perturbations; the latest Planck results give $A_s = \left(2.099 \pm 0.030\right)\times 10^{-9}$ and $n_s = 0.9649 \pm 0.0042$ \cite{Planck18Parameters}, consistent with the predictions of simple single-field slow-roll models of inflation \cite{Starobinski1979, InflaLinde1982}. These models not only reproduce the nearly scale-invariant spectrum inferred from Planck but also provide good fits to the combined Planck and BICEP2/Keck Array data \cite{PlanckInflation18, BICEPKeck, JudgmentDayInflation}, remaining compatible with several classes of single-field slow-roll scenarios characterized by different potential shapes \cite{EnciclopediaInflation}.

Despite this success, a broad class of inflationary scenarios predicts localized or oscillatory departures from this simple shape, usually referred to as Primordial Features (PF). In the context of inflation, PF are particularly important because they can encode transient departures from slow-roll dynamics \cite{SearchingFeatures1, MirandaHuModeloFeature,SearchingFeatures2}, or be generated by modified scalar-field dynamics during inflation \cite{InverseScalarDynamics}, thereby offering a window into inflationary physics beyond the information carried by the tilt and amplitude alone. Common deviations from the power-law form of $P_{\mathcal{R}}(k)$ include global logarithmic oscillations arising from non-Bunch-Davies initial conditions \cite{ModelMartin2001,ModelBozza2003,ModelMartin2003} or from axion monodromy \cite{ModelFlauger2017}, global linear oscillations predicted by boundary effective field theory models \cite{ModelJackson2013,SearchingFeatures1}, localized oscillatory features induced by a step in the inflaton potential \cite{ModelAdams2001} or in the sound speed \cite{ModelAchucarro2010,ModelInflationSpeedSound}, as well as large-scale cutoffs \cite{ModelContaldi2003,Parametric2Sinha2006}, oscillatory features from heavy fields \cite{ModelAchucarro2010, CostaHeavyFields}, and more general modulations \cite{ModelDanielsson2002,ModelChen2012}. Besides these inflationary mechanisms, primordial features in $P_{\mathcal{R}}(k)$ can also arise from non-inflationary sources such as curvaton scenarios \cite{CurvatonFeatures} and modulated reheating \cite{ModulatedReheatingFeatures}.

There are different ways to infer information about $P_{\mathcal{R}}(k)$ from cosmological data. A common approach is to adopt a parametric description, such as a power-law with $n_s$ running or a set of feature templates motivated by inflation, and then to perform parameter inference using CMB and LSS measurements \cite{ModelContaldi2003,ParametricBridle, Parametric3Simon2005,Parametric4Bridges2006,Parametric2Sinha2006,Parametric6Covi2006,Parametric5Bridges2007,Parametric7Joy2009,Parametric8Paykari2010,Parametric9Guo2011,Parametric10Goswami2013, Ballardini3, SimonySantosParametric, Ballardini2, Ballardini1, BOSSFeatures,EuclidSearchFeatures, PrimordialFeaturesEFT2025}. Alternatively, minimally parametric or non-parametric reconstructions can be explored, for example using band-power approaches, smoothing splines, sparse-recovery techniques, or knots, which aim to capture departures from a power-law \cite{ReconWaveletReal, ReconLinearInterpolation1, ReconWaveletSimulated, ReconSplinesWMAP1, ReconSplinesWMAP3, ReconCMBlike, PlanckInflation13, ReconPRISM, ReconFilter, KnotedSky, PlanckInflation15, ReconSplines, WillPaper, ReconTopHat, ReconLSSOrthogonal, DeblurringReconstruction, PlanckInflation18, MethodologicalPaper, RaffaelliKnots, JPASPaper}. These reconstruction approaches are particularly well suited to extract model-agnostic information on the shape of $P_{\mathcal{R}}(k)$ and to identify broad deviations spanning multiple scales, although very sharp or highly oscillatory signatures can remain challenging to recover within these approaches \cite{FineFeatures}.

A new generation of Stage IV LSS surveys are delivering high-precision measurements of galaxy clustering over unprecedented volumes, substantially improving our ability to test inflationary physics beyond the simplest power-law description of $P_{\mathcal{R}}(k)$ \cite{EuclidSpecifications,JPASSpecifications,DESISpecifications}. Compared to CMB-only constraints, these surveys are expected to be particularly informative at intermediate scales, where power-spectrum measurements can reach high signal-to-noise, and where physically motivated primordial modulations may be observable. Different experimental strategies contribute in complementary ways: spectroscopic surveys provide accurate three-dimensional clustering thanks to precise redshift determinations; photometric surveys observe larger samples and sky areas at the expense of radial resolution; and hybrid approaches combine both types of information. Prominent examples include the Dark Energy Spectroscopic Instrument (DESI) \cite{DESISpecifications}, which is designed as a high-density spectroscopic survey, the ESA Euclid mission \cite{EuclidSpecifications}, which combines broad-band photometry with near-infrared slitless spectroscopy, and the Javalambre Physics of the Accelerating Universe Astrophysical Survey (J-PAS) \cite{Benitez2009, JPASSpecifications}, which relies on many narrow-band filters to achieve quasi-spectroscopic redshift precision with photometric techniques. These varied capabilities motivate applying model-independent reconstruction methods to existing datasets, quantifying the current constraints that upcoming Stage IV analyses will build upon.

Although Stage IV surveys such as DESI and Euclid will ultimately deliver substantially tighter constraints on inflationary physics from galaxy clustering, their cosmological datasets are still in the process of being built through successive public releases. For instance, DESI has already released its DR1 based on the first 13 months of the main survey campaign \cite{DESIDR1}, and Euclid is following a phased release strategy in which early and quick releases precede the main data releases as the survey progresses.\footnote{According to the official Euclid data release timeline \\(\mbox{\url{https://euclid.caltech.edu/page/data-release-timeline}}), the first main data release (DR1) is expected to be released in the second half of 2026.} \cite{EuclidQ1}. As a result, even when Stage IV data are available, analyses that aim to be stable and end-to-end are best developed and tested on completed catalogs. These stage III surveys provide mature, final datasets with well-tested clustering pipelines, enabling a clean methodological validation and a robust pre-Stage-IV baseline for primordial-feature searches \cite{DESGeneric1, BOSSGeneral1,BOSSGeneral2, DESGeneric2, SDSSDR16}. In this work we use clustering measurements from the final Baryon Oscillation Spectroscopic Survey (BOSS) Data Release 12 (DR 12) \cite{BOSSDR11and12, BOSSDR12Catalogue} and extended Baryon Oscillation Spectroscopic Survey (eBOSS) Data Release 16 (DR 16) \cite{eBOSSDR16, eBOSS2} to reconstruct $P_{\mathcal{R}}(k)$ over the range of comoving wavenumbers where these surveys retain constraining power, and to search for statistically significant deviations from the power-law spectrum. We investigate the robustness of the reconstruction under different analysis choices, including redshift binning and galactic hemisphere, and, when appropriate, combine the available information to obtain the tightest constraints achievable with BOSS and eBOSS.

The structure of the paper is as follows: in \cref{sec:BOSS} we describe the BOSS and eBOSS data products and the mock catalogs used for validation; in \cref{sec:Methodology} we present the $P_{\mathcal{R}}(k)$ knot reconstruction methodology; in \cref{sec:GalaxyPowerSpectrumModel} we detail the parametric modeling needed to connect the primordial power spectrum to the measured galaxy power spectrum; in \cref{sec:Validation} we validate the methodology and its application to BOSS/eBOSS, also discussing the set of priors to be used; in \cref{sec:Results} we report the results obtained from the data; and in \cref{sec:Conclusions} we present the conclusions derived from this work and possible lines of future work. In Appendix~\ref{subsec:AppendixBAO} we provide the BAO smoothing technique used; in Appendix~\ref{subsec:RobustnessNs} we assess the robustness of the $n_s$ that we infer; and in Appendix~\ref{subsec:PLPriors} we perform the $P_{\mathcal{R}}(k)$ reconstruction with an additional prior set.

\section{The BOSS/eBOSS surveys}
\label{sec:BOSS}
 
The Sloan Digital Sky Survey (SDSS) Baryon Oscillation Spectroscopic Survey (BOSS) \cite{SDSSIII,BOSSDR12Catalogue} and its successor, the extended Baryon Oscillation Spectroscopic Survey (eBOSS) \cite{SDSSIV,eBOSS2}, constitute two landmark spectroscopic programmes for LSS studies. Their combined legacy includes precise measurements of Baryon Acoustic Oscillations (BAO) and Redshift-Space Distortions (RSD) \cite{BOSSDR12BAOandRSD,BOSSDR12BAOandRSD2}, which provide complementary information on the distance--redshift relation and the growth of cosmic structure. Beyond late-time cosmology, the large volumes and broad range of comoving scales probed by BOSS and eBOSS enable sensitive tests of the primordial power spectrum \cite{BOSSPNG,BOSSDataBeutlerInspiration}. The presence of PF can imprint characteristic modulations in Fourier-space clustering observables, motivating searches using power spectrum measurements.

In this work, we use clustering measurements from the final BOSS and eBOSS data releases\footnote{\url{https://fbeutler.github.io/hub/deconv_paper.html}.} \cite{BOSSDataBeutlerInspiration, BOSSDataBeutler}. In the remainder of this section we summarize the main survey characteristics relevant to our analysis and, subsequently, describe the galaxy power spectrum measurements and mock catalogs employed in the methodological validation.

\subsection{Survey specifications}
\label{subsec:BOSSSpecifications}

BOSS obtained spectroscopic redshifts for $1{,}198{,}006$ luminous red galaxies (LRG) over a sky area of $10{,}252~\mathrm{deg}^2$, spanning the redshift range $0.2<z<0.75$. The footprint is conventionally split into two disjoint regions, the North Galactic Cap (NGC) and the South Galactic Cap (SGC), which can be analyzed separately to perform internal consistency tests and to mitigate potential observational systematics that differ between hemispheres. The BOSS DR 12 LRG sample is commonly analyzed in two broad, independent redshift bins, $0.2<z<0.5$ ($z_1$) and $0.5<z<0.75$ ($z_3)$, with effective redshifts $z_{\rm eff} = 0.38$ and $z_{\rm eff} = 0.61$, respectively. In this work, we do not employ the intermediate redshift bin $0.4<z<0.6$ ($z_2$), as it completely overlaps with $z_1$ and $z_3$. For simplicity, we only consider the non-overlapping $z_1$ and $z_3$ bins, and assume them to be independent, as is done in BOSS official analyses. 

The eBOSS programme extended SDSS spectroscopic clustering measurements to higher redshift by targeting multiple tracers, including high-redshift LRG, emission-line galaxies (ELG), and quasars (QSO). In the final eBOSS catalogs, the quasar sample provides redshifts over $0.8<z<2.2$, while the LRG and ELG samples probe roughly $0.6<z<1.0$ and $0.6<z<1.1$, respectively. The final eBOSS quasar catalog contains $\simeq 3.4\times 10^5$ objects over  $4808~\mathrm{deg}^2$, and the eBOSS LRG sample includes $174{,}816$ redshifts over $4242~\mathrm{deg}^2$ in $0.6<z<1.0$ \cite{eBOSS2}. For simplicity, we consider only the eBOSS QSO, excluding the LRG and ELG samples. We adopt this approach because the eBOSS LRG are correlated with the BOSS $z_3$ LRG, which would increase the computational cost of our analysis by requiring the construction of a combined covariance matrix. In addition, the eBOSS ELG sample requires a complex treatment of systematics that, if not properly modeled, could introduce spurious large-scale effects.

Taken together, BOSS LRG and eBOSS QSO provide sensitivity to very large comoving scales, and span a wide redshift range. These characteristics are particularly advantageous for searches of primordial features that may produce scale-dependent modulations in the measured power spectrum.

\subsection{Power spectrum data}
\label{subsec:BOSSData}

We adopt the redshift binning used in \cite{BOSSDataBeutler}, splitting the BOSS LRG sample into two non-overlapping bins, $z_1$ and $z_3$, while keeping all the eBOSS QSO in a single redshift bin. The properties of these different bins are reported in \cref{tab:BOSSBins}.
\begin{table}[t]
  \centering
  \begin{tabular}{|l c c c c c|}
\hline
Survey/Sample & $N_{\rm gal}$ & $[z_{\min},z_{\max}]$ & $z_{\rm eff}$ & $V_{\rm eff}\,[{\rm Gpc}^3]$ & \# mocks \\
\hline\hline
BOSS DR12 LRG NGC $z_1$ & 429\,182 & [0.2, 0.5] & 0.38 & 2.6 & 2048 \\
BOSS DR12 LRG SGC $z_1$ & 174\,820 & [0.2, 0.5] & 0.38 & 1.0 & 2048 \\
BOSS DR12 LRG NGC $z_3$ & 435\,742 & [0.5, 0.75] & 0.61 & 2.8 & 2048 \\
BOSS DR12 LRG SGC $z_3$ & 158\,262 & [0.5, 0.75] & 0.61 & 1.0 & 2048 \\
eBOSS DR16 QSO NGC & 218\,209 & [0.8, 2.2] & 1.52 & 0.35 & 1000 \\
eBOSS DR16 QSO SGC & 125\,499 & [0.8, 2.2] & 1.52 & 0.18 & 1000 \\
\hline
\end{tabular}

    \caption{Properties of the different BOSS/eBOSS samples studied in this paper. $N_{\text{gal}}$ is the number of galaxies, $z_{\text{eff}}$ the effective redshift, and $V_{\text{eff}}$ the effective volume. We include the number of available mock realizations for each bin.}
  \label{tab:BOSSBins}
\end{table}

Both BOSS LRG DR 12 and eBOSS QSO DR 16 power spectra, $\hat{\text{P}}_g$, are provided in an array of 400 linearly spaced $k$-bins, $\hat{\text{k}}$, covering the range $0.0018 \leq k \leq 0.4\text{ h}\,\mathrm{Mpc}^{-1}$ for each $z$-bin. In order to compute the number of modes in each bin, we create upper and lower scale bin arrays, $\hat{\text{k}}_{\text{\text{sup}}}$ and $\hat{\text{k}}_{\text{\text{inf}}}$, which are $\hat{\text{k}}+0.001 \text{ h}\,\mathrm{Mpc}^{-1}$ and $\hat{\text{k}}-0.001 \text{ h}\,\mathrm{Mpc}^{-1}$ respectively\footnote{Except for the first element of $\hat{\text{k}}_{\text{\text{inf}}}$, which is set to 0.}. Then, we calculate the number of modes per volume unit for each scale of $\hat{k}$ as: 
\begin{equation}\label{NumberOfNodes}
N_k(\hat{k}) =
\frac{\frac{4}{3}\pi\left(\hat{k}_{\mathrm{sup}}^{3}-\hat{k}_{\mathrm{inf}}^{3}\right)}{(2\pi)^3}.
\end{equation}
To reduce computational cost, we rebin the data into 40 scale-bins. We recalculate the $k$-bins and the monopole galaxy power spectra according to the number of modes $N_k(k)$, obtaining new arrays $\text{k}$ and $\text{P}_g^{\text{(0)}}$, whose components are:
\begin{align}
k_j =
\frac{\sum_{i=10(j-1)+1}^{10j} N_k(\hat{k}_i)\,\hat{k}_i}
{\sum_{i=10(j-1)+1}^{10j} N_k(\hat{k}_i)},
\qquad
P^{(0)}_{g,j} =
\frac{\sum_{i=10(j-1)+1}^{10j} N_k(\hat{k}_i)\,\hat{P}^{(0)}_{g,i}}
{\sum_{i=10(j-1)+1}^{10j} N_k(\hat{k}_i)}.
\label{P040Bins}
\end{align}

The covariance matrix $\text{C}$ is given in these 40 scales. With all those ingredients, the likelihood $\mathcal{L}$ can be constructed by comparing the data with a model, which will be detailed below:
\begin{equation}\label{eq:Likelihood}
\mathcal{L} \propto \exp\left[
-\frac{1}{2}
\left( P_{g}^{\text{(0)}}(\text{data}) - \text{W} P_{g}^{\text{(0)}}(\text{model})  \right)^{T}
C^{-1}
\left(  P_{g}^{\text{(0)}}(\text{data}) - \text{W} P_{g}^{\text{(0)}}(\text{model})\right)
\right]
\end{equation}
where $W$ accounts for the selection function (window) effects. 

\subsection{Mock catalogs}
\label{subsec:BOSSMocks}

Mock catalogs play a central role in clustering analyses. On the one hand, they provide an end-to-end validation that allows the identification of potential systematics in the employed pipelines. On the other hand, when a sufficiently large number of realizations is available, they are also used to numerically estimate the covariance matrix entering the likelihood of \cref{eq:Likelihood}. 

\begin{itemize}
\item The \textsc{MD-Patchy} mocks \cite{MDPatchyMocks} employ the Augmented Lagrangian Perturbation Theory prescription \cite{ModellingBOSSPerturbationTheory} to efficiently generate large ensembles of synthetic galaxy catalogs that mimic the properties of full $N$-body simulations. In order to ensure such realism, these mocks are calibrated against a reference sample drawn from the high-resolution BigMultiDark simulations \cite{MultiDarkSimulations}. The BigMultiDark simulation was run with \textsc{Gadget}-2 using $3840^3$ particles in a volume of $(2.5~\mathrm{Gpc}/\text{h})^3$, assuming a $\Lambda$CDM cosmology with $\Omega_m=0.307115$, $\Omega_b=0.048206$, $\sigma_8=0.8288$, $n_s=0.9611$, and $h=0.6777$. Halo abundance matching was then applied so that the mocks reproduced the observed two- and three-point clustering of BOSS galaxies, with a redshift-dependent implementation designed to capture the survey evolution. A total of 2048 independent \textsc{MD-Patchy} mock realizations were produced describing the BOSS LRG DR 12 geometry for both NGC and SGC, over the full redshift range $0.2<z<0.75$. 

\item The \textsc{Nseries} mocks~\cite{NSeriesMocks} consist of full $N$-body realizations populated with a fixed halo occupation distribution model matching the NGC CMASS BOSS LRG DR 12 sample\footnote{The CMASS geometry corresponds to the redshift range $0.43 < z < 0.70$ ($z_{\rm eff} = 0.56$), which was commonly employed in BOSS data releases prior to DR 12.}. These mocks were generated from seven independent periodic $N$-body boxes (whose cosmology is consistent with the WMAP cosmology) of side length $2.6~\mathrm{Gpc}/\text{h}$. For each cubic box, four different orientations of the CMASS northern geometry are fitted, and each of these orientations is applied to three pre-rotation positions of the box, where the three Cartesian positions and velocities are swapped. This procedure yields a total of 84 pseudo-independent realizations with the NGC CMASS BOSS geometry. Although these 84 realizations do not represent a sufficiently large sample to estimate a reliable covariance matrix, these mocks are essential for testing potential systematics in the analysis pipeline, as they accurately represent the full $N$-body realism. 

\item The \textsc{EZmock} catalogs~\cite{eBOSSEZMocks} comprise realizations generated with the Extended Zel'dovich (EZ) \cite{EZMocks} approximate $N$-body scheme, with free parameters calibrated to reproduce the desired two- and three-point clustering. Redshift evolution is included by constructing a light cone from seven redshift shells drawn from periodic boxes of side length $L = 5000\text{ h}^{-1}\mathrm{Mpc}$. The underlying cosmology of these mocks is the same as that used for the \textsc{MD-Patchy}  mocks. The eBOSS DR 16 quasar geometry is imprinted, with a total of 1000 independent realizations for both the NGC and SGC regions.  
\end{itemize}

\section{Primordial power spectrum reconstruction}
\label{sec:Methodology}

We adopt a model-independent method to reconstruct the primordial power spectrum $P_{\mathcal{R}}(k)$. This is achieved by freely sampling $N$ knots in the $\log\{k, P_{\mathcal{R}}\}$ plane, using the nested sampler \textsc{PolyChord} \cite{MandatoryPolyChord1,MandatoryPolyChord2}. A detailed description of this methodology can be found in \cite{MethodologicalPaper}, while a very brief summary is provided in this section.

\textsc{Cobaya} \cite{CobayaMandatory2,CobayaMandatory1} is used to implement both the model and the likelihood for the $P_{\mathcal{R}}(k)$ reconstructions, as well as to run the nested sampler \textsc{PolyChord}. We explore a knot number $N$ from 2 to 7, and subsequently marginalize over this parameter. For a knot configuration with $N$ knots, the maximum number of sampled parameters is $2N + 5$: $2N - 2$ knot parameters $\Theta_{\text{knots}}$ (since the outermost knots are scale-fixed), and seven galaxy/quasar power spectrum model parameters $\Theta_{\text{model}}$ per redshift bin, as will be detailed in the next section. \textsc{PolyChord} is executed with $25$ live points per sampled parameter, with $\texttt{num\_repeats\_value} = 24 N$, and with a stopping tolerance given by the parameter \texttt{precision\_criterion} set to $10^{-3}$. The numerical uncertainty on the log-evidence reported by PolyChord is typically below $0.5\%$ in all cases. The reconstructions presented in this work required about $3 \times 10^4$ CPU hours, using $128$ CPU cores for each sampling case. The employed priors will be discussed in \cref{sec:Validation}.

For a given $P_{\mathcal{R}}(k)$ reconstruction, the evidences $Z_N$ are also estimated with \textsc{PolyChord}. From them,
we compute the Bayes factor $Z_2 / Z_{\text{max}}$, where $Z_{\text{max}}$ denotes the highest evidence among all $Z_N$ values excluding $Z_2$. Configurations with $N > 7$ do not substantially contribute to the reconstructions, since $Z_N$ display a decreasing trend with $N$, and above $N = 7$ configurations yield evidences below the $1\%$ contribution in all cases. Our non-parametric $P_{\mathcal{R}}(k)$ reconstruction is obtained combining the reconstructions from all $N$-knot configurations into a posterior that is effectively marginalized over $N$. This posterior is computed from an evidence-weighted averaging.

As with previous applications of this methodology, the sampled coordinates are ${x,y}$ for the reconstructions: $x$ corresponds to $\log k$ (from 0 to 1, according to the survey scale grid), and $y$ to $\log (P_{\mathcal{R}}(k))$. These coordinates are sorted in scale, as required to solve the label switching problem \cite{MethodologicalPaper}.

\section{Galaxy power spectrum}
\label{sec:GalaxyPowerSpectrumModel}

In this section, we show that the primordial feature signal can be propagated into the evolved galaxy field at linear order. PF signals cannot be reproduced by any particular combination of cosmological parameters within the Standard Model. Assuming that we can marginalize over the broadband shape of the galaxy power spectrum, then we can search for residual feature signals in galaxy clustering. We describe in this section the galaxy power spectrum parametric model that we use to reconstruct $P_{\mathcal{R}}(k)$ in this work.

The Standard Model power-law is:
	\begin{equation}\label{PowerLawPPS}
P_{\mathcal{R},0}(k) = A_s \left( \frac{k}{k_0} \right)^{n_s - 1}.
	\end{equation}
We consider potential deviations of the power-law in the form of features. The contributions of primordial features, $\delta P_{\mathcal{R}}(k)$, appear as corrections of the power-law:
	\begin{equation}\label{FeaturesPPS}
P_{\mathcal{R}}(k) = P_{\mathcal{R},0}(k) [1 + \delta P_{\mathcal{R}}(k)].
	\end{equation}
   
The linear matter power spectrum $P_m(k)$ is derived from $P_{\mathcal{R}}(k)$ by means of the transfer function $T(k)$ and the growth function $D(z)$:
\begin{equation}\label{MatterPS}
    P_m(k,z) = T^2(k)\,D^2(z)\,P_{\mathcal{R}}(k).
\end{equation}
Non-linear matter clustering causes the smoothing of the PF and BAO signals. To account for this effect, it is useful to isolate the BAO and PF signals from the broadband. We need to construct a no-wiggle matter spectrum $P_{\text{nw}}(k)$, which is obtained by smoothing the BAO wiggles from the matter power spectrum $P_m(k)$. In Appendix~\ref{subsec:AppendixBAO}, we provide a description of the employed modeling for the BAO smoothing to compute $P_{\rm nw}(k)$. Consequently, we can write the matter power spectrum as: 
	\begin{equation}\label{MatterDecomposedPS}
P_m(k) =  P_{\text{nw}}(k)  \left[1 + \mathcal{O}_{\text{lin}}(k)  + \delta P_{\mathcal{R}}(k) + \mathcal{O}_{\text{lin}}(k) \delta P_{\mathcal{R} }(k) \right],
    \end{equation}
where $\mathcal{O}_{\text{lin}}(k) \equiv \frac{P_m(k)}{P_{\text{nw}}(k)} - 1$ is defined for the featureless case, $\delta P_{\mathcal{R}}(k)=0$.

\Cref{MatterDecomposedPS} shows the result for the matter field, which must be related to the observable quantity, the galaxy power spectrum. Since we focus solely on modulations around the smooth power spectrum, we adopt a phenomenological parametric model rather than the standard perturbation theory bias expansion. In this approach, the relation between the matter and galaxy fields is described by broadband information encoded in a linear bias $b$, an effective angle-averaged spectral distortion from the non-linear velocity field, $F(k, \Sigma_s)$, and a set of broadband parameters, $A(k)$ \cite{BOSSFeatures}. We can write the monopole component as:
	\begin{equation}\label{GPS}
P^{(0)}_{g,\text{nw}}(k) = b^2 P_{\text{nw}}(k) \, F(k,\Sigma_s) + A(k),
	\end{equation}
where the functions $A(k)$ and $F(k, \Sigma_s)$ are defined as:
\begin{equation}\label{FFunction}
F(k, \Sigma_s) = \frac{1}{\left(1 + k^2 \Sigma_s^2 / 2\right)^2},
\end{equation}
and
\begin{equation}\label{AFunction}
A(k) = \frac{a_1}{k^2} + \frac{a_2}{k} + a_3.
\end{equation}
The functional forms of $A(k)$ and $F(k, \Sigma_s)$ are well motivated and widely adopted in clustering analyses \cite{AandFReference1,AandFReference2,AandFReference4, AandFReference3,AandFReference5}.

The late-time gravitational evolution driven by bulk flows smooths features in the linear power spectrum—such as the BAO and PF—so we model this effect with a damping parameter, $\Sigma_{\rm nl}$. Infrared-resummation and $N$-body studies show that large-scale displacements act on the total density field and damp generic structures in $P_m(k)$ through an approximately Gaussian (Zel'dovich-like) propagator, which justifies using the same damping for both BAO wiggles and the PF. Therefore, the monopole galaxy power spectrum is:
\begin{equation}\label{Pdewiggled}
P_g^{\text{(0)}}(k) = P_{g,\text{nw}}^{(0)}(k) \left[1 + \left[ O_{\text{lin}}(k) + \delta P_{\mathcal{R}}(k) + O_{\text{lin}}(k) \delta P_{\mathcal{R} }(k) \right] \text{e}^{-k^2 \Sigma_{\rm nl}^2/2} \right]
\end{equation}

An additional aspect to consider is the inclusion of the isotropic BAO scaling parameter, $\alpha$. In Fourier-space analyses, this parameter determines the characteristic scale of the BAO. Thus, the full model employed in this analysis is:
\begin{equation}\label{PModel}
P_g^{\text{(0)}}(k) = P_{g,\text{nw}}^{(0)}(k) \left[1 + \left[ O_{\text{lin}}(k/\alpha) + \delta P_{\mathcal{R}}(k) + O_{\text{lin}}(k/\alpha) \delta P_{\mathcal{R} }(k) \right] \text{e}^{-k^2 \Sigma_{\rm nl}^2/2} \right].
\end{equation}
In total, 7 model parameters are needed: $\Theta_{\text{model}} = \{b, \Sigma_s, \Sigma_{\rm nl}, a_1, a_2, a_3, \alpha\}$. \sloppy Each BOSS/eBOSS $z$-bin and galactic cap will have an independent set of model parameters (including $\alpha$), while the $\Theta_{\text{knots}}$ that generate the $P_{\mathcal{R}}(k)$ reconstruction are the same for all bins and galactic caps. 

We reconstruct $P_{\mathcal{R}}(k)$ over the scale range $k \in [0.007, 0.3] \text{ h} \text{ Mpc}^{-1}$, divided into 30 $k$-bins with $\Delta \log k = 0.055$. The lower bound of $0.007 \text{ h} \text{ Mpc}^{-1}$ corresponds to the largest scale at which BOSS retains a sufficient signal-to-noise ratio, while the upper bound of $0.3 \text{ h} \text{ Mpc}^{-1}$ is limited by the Nyquist frequency. Our $P_g^{\text{(0)}}(\text{model})$ needs to be computed up to $0.4 \text{ h} \text{ Mpc}^{-1}$, since the window function $W(k)$ is informative at scales $k \in [0.3,0.4] \text{ h} \text{ Mpc}^{-1}$, impacting our $[0.007,0.3] \text{ h} \text{ Mpc}^{-1}$ range. For computing $\delta P_{\mathcal{R}}(k)$, we take as a power-law reference a linear interpolation between the outermost knots\footnote{A different power-law baseline was tested, constructed from a least-squares fit to all the sampled knots. The impact on both the evidences and the reconstruction contours was small in comparison with the outer-knot baseline used throughout the analysis.}. Thus, the $N = 2$ configuration represents a featureless scenario by definition. From this power-law reference, we construct a fiducial fixed-template linear matter power spectrum, $P_m(k)$, using the Planck DR3 cosmological parameters $\Theta_{\text{cosmology}} = \{H_0,\Omega_c h^2,\Omega_b h^2\}$. We then account, without loss of generality, for variations in the smooth galaxy power spectrum $P_{g,\mathrm{nw}}(k)$ through changes in $\Theta_{\text{model}}$ and $\Theta_{\text{knots}}$. By following the standard BAO approach, we fix the cosmological template $P_m(k)$ to the Planck DR3 parameters ($\Theta_{\rm cosmology}$), while the galaxy model parameters $\Theta_{\rm model}$ and the reconstruction knots $\Theta_{\rm knots}$ marginalize over the broadband shape of the galaxy power spectrum. This allows us to isolate residual primordial feature signals that cannot be reproduced by any combination of Standard Model cosmological parameters.

The likelihood function $\mathcal{L}$ is evaluated by comparing the model galaxy power spectra, $P^{(0)}_{g}(\mathrm{model})$, with those derived from BOSS data, $P^{(0)}_{g}(\mathrm{data})$. This comparison is made through all the predefined $k$-grid, across multiple redshift bins and galactic caps. The likelihood is modeled as a multivariate Gaussian:
\begin{align}
-2 \log(\mathcal{L}) 
&= \sum_{\gamma} \sum_{\mu} 
\left[\boldsymbol{P}^{(0)}_{g,\{\gamma,\mu\}}(\mathrm{model}) 
- \boldsymbol{P}^{(0)}_{g,\{\gamma,\mu\}}(\mathrm{data})\right]^{T} \notag \\
&\quad \textbf{C}_{\{\gamma,\mu\}}^{-1}
\left[\boldsymbol{P}^{(0)}_{g,\{\gamma,\mu\}}(\mathrm{model}) 
- \boldsymbol{P}^{(0)}_{g,\{\gamma,\mu\}}(\mathrm{data})\right],
\label{likelihood}
\end{align}
where $\gamma$ indexes the galactic caps and $\mu$ the redshift bins. For each pair $\{\gamma,\mu\}$, the quantities $\boldsymbol{P}^{(0)}_{g,\{\gamma,\mu\}}(\mathrm{model})$ and $\boldsymbol{P}^{(0)}_{g,\{\gamma,\mu\}}(\mathrm{data})$ are vectors with components defined over the sampled $k$-bins. $\mathrm{\textbf{C}}_{\{\gamma,\mu\}}$ denotes the corresponding BOSS or eBOSS covariance matrices, obtained from the \textsc{MD-Patchy} or \textsc{EZmocks}, respectively.



\section{Methodological validation}
\label{sec:Validation}

Before reconstructing the $P_{\mathcal{R}}(k)$ directly from the BOSS+eBOSS data, we first validate our methodology. Although it has been tested previously \cite{MethodologicalPaper} and used in a J-PAS forecast \cite{JPASPaper}, in both cases it was applied to galaxy power spectrum linear models with a physically motivated description of RSD. Thus, we begin by assessing the performance of the methodology applied to the parametric galaxy power spectrum model described in the previous section, testing the primordial feature sensitivity. As last validation step, we apply the $P_{\mathcal{R}}(k)$ reconstruction to the BOSS/eBOSS mock catalogs.

\subsection{Validation of primordial feature detection capabilities}
\label{subsec:ValidationPFModel}

We assess the sensitivity to PF of the $P_{\mathcal{R}}(k)$ reconstructions using the parametric galaxy power spectrum model described in \cref{PModel}. Since that model is defined with respect to a fixed template cosmology, all $P_{\mathcal{R}}(k)$ reconstructions in this work are performed fixing the $\Theta_{\text{cosmology}}$ of Planck DR3. For this parametric model, part of the cosmological dependence can be effectively absorbed by phenomenological parameters such as $\alpha$ and the broadband coefficients $a_i$. To test the PF sensitivity, we construct a realization in which the input $P_{\mathcal{R}}(k)$ is controlled, and we assume either a power-law primordial spectrum or Local Oscillatory (LO) features with 10\%, 5\%, or 3\% amplitudes. These realizations are generated by fixing $\Theta_{\text{model}}$ to the median value of the best-fit of the mocks, with the same survey window function and covariance matrix as with the data. Therefore, the only difference among the different realizations is their primordial power spectrum.

First, we impose uniform priors on both $\Theta_{\text{knots}}$ and $\Theta_{\text{model}}$ (except the bias, which is fixed due to its degeneracy with the amplitude). Although the power-law case is properly recovered, the presence of the 10\% local oscillatory feature is not detected, as the strong degeneracies between $\Theta_{\text{model}}$ and $\Theta_{\text{knots}}$ hinder its recovery. Moreover, the reconstruction contours remain very wide, which we aim to improve.

\begin{figure}[t]
\centering 
\includegraphics[width=.55\textwidth]{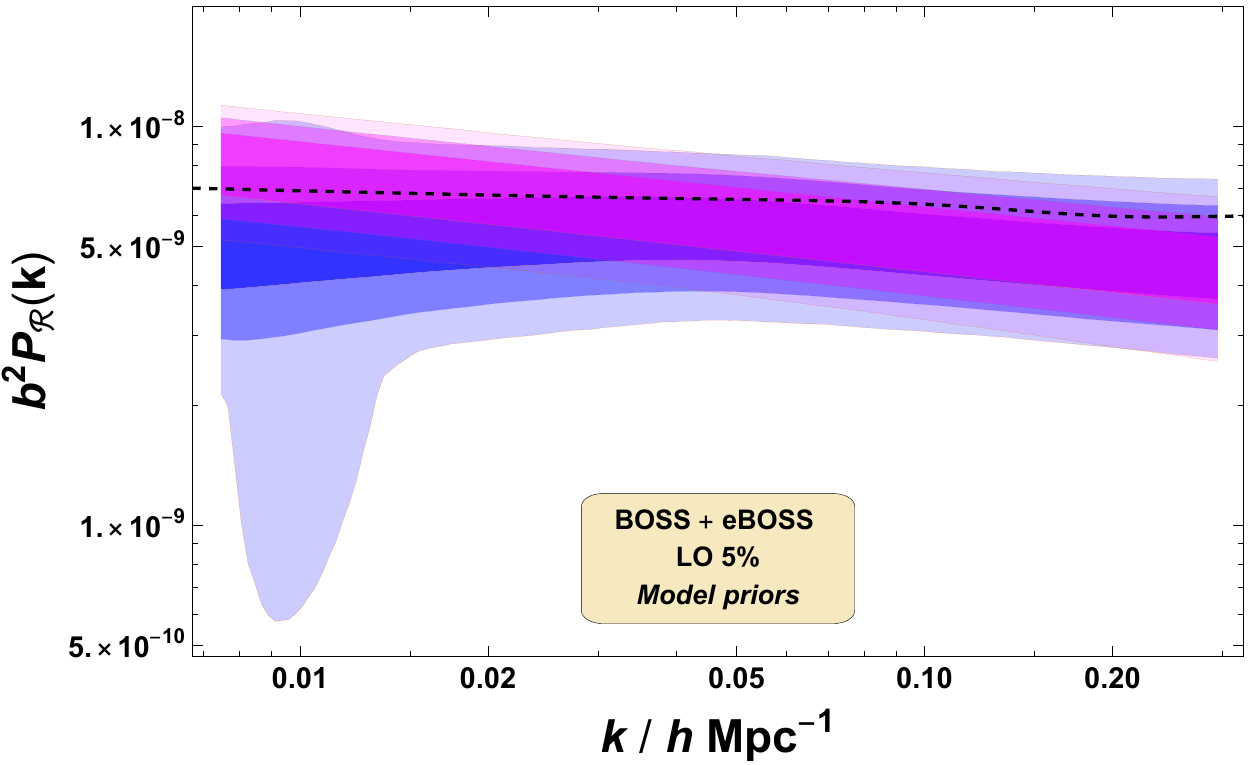}
\includegraphics[width=.39\textwidth]{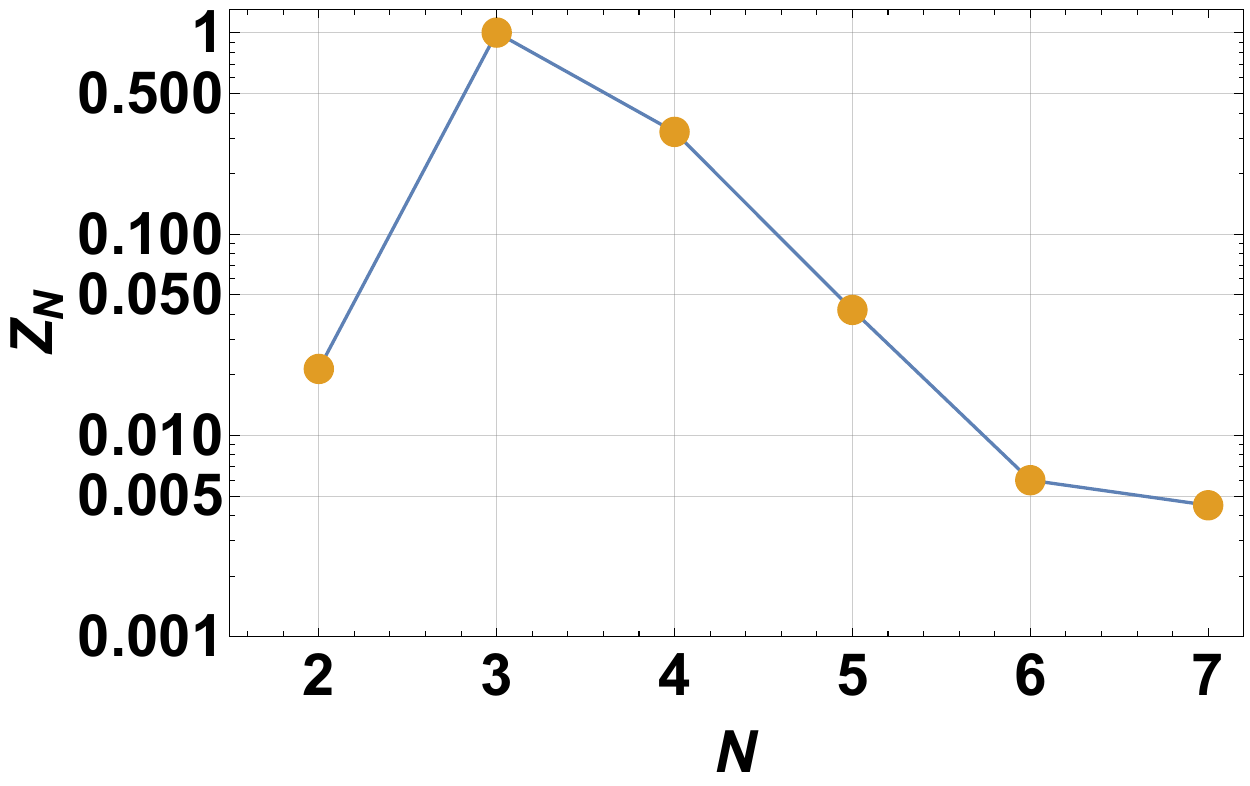}
\caption{Left: $P_{\mathcal{R}}(k)$ reconstruction for a BOSS + eBOSS realization with a 5\% input local oscillatory feature (dashed black line), using the \textit{model priors}. Magenta contours represent the power-law–equivalent two-knot reconstruction, and blue contours the $N$-marginalized ones. Since the bias is non-informatively sampled, the amplitude of the spectrum is completely degenerate with $b^2$. Therefore, $b^2 P_{\mathcal{R}}(k)$ is shown in the vertical axis. Right: the evidences $Z$ as function of the number of knots $N$.}
\label{fig:RealizationBestFitMock}
\end{figure}

In order to be able to spot LO features and to improve reconstruction precision, we impose a set of informative priors on $\Theta_{\text{model}}$, derived from the mock catalogs described in \cref{subsec:BOSSMocks}. These new priors are referred to as \textit{model priors}, for which informative Gaussian priors are imposed on the model parameters $\Theta_{\text{model}}$, while the knot parameters are sampled from uniform priors. In the next subsection, we describe how to infer the numerical values of this prior set, collected in \cref{tab:PriorsSpecific}. We perform the $P_{\mathcal{R}}(k)$ reconstruction for both BOSS and eBOSS catalogs with the \textit{model priors}: the degeneracies are mitigated and the recovery of a 10\% local oscillatory feature is achieved with decisive statistical evidence. We also reconstruct the 5\% local oscillatory feature, displayed in \cref{fig:RealizationBestFitMock}. In this case of a smaller-amplitude feature, we obtain very strong evidence for the feature detection, with the $3$ and $4$-knot configurations providing a higher evidence than the two-knot case. If we reconstruct for the LO 3\% or the power-law templates, we obtain no evidence of PF.

\subsection{Methodological validation with mocks}
\label{subsec:MockValidation}

Once the PF sensitivity has been validated, we use the BOSS \textsc{MD-Patchy} and eBOSS \textsc{EZmocks} catalogs to derive the values of the model parameters $\Theta_{\text{model}}$ that constitute the \textit{model priors}.

While keeping the fiducial $\Theta_{\text{cosmology}}$ of the mocks fixed, we sample $\Theta_{\text{model}}$ adopting uniform priors. With the resulting posteriors, we estimate the mean best-fit values $\bar{\Theta}_{\text{model}}$ and the covariance matrices $\bar{C}_{\text{model}}$.
Each chain $n$ for a given $\Theta_{\text{model}}$ derived from a mock $m$ is combined as:
\begin{equation}\label{MeanComputation}
\bar{\Theta}_{\text{model}} =
\sum_{m=1}^{M} \sum_{n} Z^m w_n^m \, \bar{\Theta}_{n}^m
\end{equation}
\begin{equation}\label{CovComputation}
\bar{C}_{\text{model}} =
\sum_{m=1}^{M} \sum_{n} Z^m w_n^m \,
\left( \bar{\Theta}_{n}^m - \bar{\Theta}_{\text{model}} \right)
\left( \bar{\Theta}_{n}^m - \bar{\Theta}_{\text{model}} \right)^{\mathrm{T}},
\end{equation}
with $Z^m$ the evidence for the mock $m$, $\omega^m_n$ the importance weight of its $n$ chain, and $\Theta^m_n$ the sampled set of model parameters. We use $M = 2000,84$, and $890$ for the \textsc{MD-Patchy}, \textsc{Nseries} and \textsc{EZmocks} respectively. We summarize in \cref{tab:BestFitParametersMocks} the mean best-fit parameters obtained from the mock catalogs. The results are compatible between all the $z$-bins and hemispheres, with the \textsc{Nseries} mocks providing tighter constraints on $\alpha$ and $\Sigma_{\mathrm{nl}}$. The mean value of the \textsc{Nseries} $\Sigma_{\mathrm{nl}}$ is about a $30\%$ smaller than for the \textsc{MD-Patchy} mocks, as expected \cite{BOSSFeatures} \cite{AandFReference3}. Nevertheless, we construct the \textit{model priors} from the \textsc{MD-Patchy} and \textsc{EZmocks} mocks, since we have better statistics with these mocks ($\approx 50$ times more mocks available than \textsc{Nseries}), and since the covariance matrix of the data has been estimated with these mocks. In particular, the Gaussian prior adopted for $\Sigma_{\text{nl}}$ is already broad enough to encompass the more realistic damping level found by the \textsc{Nseries} mocks. All these mean best-fit values and their variances have successfully passed a $\chi^2$ test, which allows us to construct the \textit{model priors} adopting correlated multivariate Gaussian priors centered on $\bar{\Theta}_{\text{model}}$ with correlations according to $\bar{C}_{\text{model}}$. These \textit{model priors} were already used to assess the PF sensitivity. In \cref{tab:PriorsSpecific} we summarize the numerical values of all the used priors.
\begin{table}[t]
    \centering
    \footnotesize
    \setlength{\tabcolsep}{0.5pt}
    \renewcommand{\arraystretch}{1.05}
    \begin{tabular}{|c|c|c|c|c|c|c|c|}
\hline
\multicolumn{1}{|c|}{\textbf{Catalog}} &
\multicolumn{5}{c|}{\textbf{BOSS galaxies}} &
\multicolumn{2}{c|}{\textbf{eBOSS QSOs}} \\
\hline
\textbf{Mock} 
& \multicolumn{4}{c|}{\textbf{MD-Patchy}} 
& \textbf{NSeries}
& \multicolumn{2}{c|}{\textbf{EZ mocks}} \\
\hline
\textbf{$\bar{\Theta}_{\text{model}}$} 
& \textbf{NGC $z_1$} 
& \textbf{NGC $z_3$} 
& \textbf{SGC $z_1$} 
& \textbf{SGC $z_3$} 
& \textbf{NGC} CMASS
& \textbf{NGC}
& \textbf{SGC} \\
\hline
$b$ 
& $1.70 \pm 0.23$ 
& $1.83 \pm 0.31$ 
& $1.55 \pm 0.31$ 
& $1.65 \pm 0.35$ 
& $1.64 \pm 0.24$
& $2.59 \pm 0.56$
& $2.27 \pm 0.63$ \\
$\Sigma_{\text{s}} [{\rm Mpc}\text{ h}^{-1}]$ 
& $2.5 \pm 1.7$ 
& $2.9 \pm 2.0$ 
& $3.9 \pm 2.6$ 
& $4.0 \pm 2.8$ 
& $3.7 \pm 2.0$
& $3.0 \pm 2.1$
& $4.2 \pm 2.6$ \\
$\Sigma_{\text{nl}} [{\rm Mpc}\text{ h}^{-1}]$ 
& $12 \pm 3.5$ 
& $11 \pm 3.4$ 
& $13 \pm 4.7$ 
& $13 \pm 4.7$ 
& $7.9 \pm 2.1$
& $15.4 \pm 8.2$
& $15.4 \pm 8.3$ \\
$a_1 [{\rm Mpc}\text{ h}^{-1}]$ 
& $-4.5 \pm 3.7$ 
& $-5.4 \pm 3.6$ 
& $-6.9 \pm 5.3$ 
& $-6.1 \pm 4.6$ 
& $-5.86 \pm 3.40$
& $-0.21 \pm 3.7$
& $-2.1 \pm 4.1$ \\
$10^{-3}a_2  [({\rm Mpc}\text{ h}^{-1})^2]$ 
& $0.82 \pm 0.47$ 
& $0.88 \pm 0.53$ 
& $1.2 \pm 0.61$ 
& $1.2 \pm 0.59$ 
& $1.00 \pm 0.43$
& $-0.04 \pm 0.6$
& $0.31 \pm 0.64$ \\
$10^{-3}a_3  [({\rm Mpc}\text{ h}^{-1})^3]$ 
& $-0.57 \pm 0.83$ 
& $-0.63 \pm 0.97$ 
& $-1.2 \pm 1.2$ 
& $-1.3 \pm 1.2$ 
& $-0.75 \pm 0.83$
& $-0.14 \pm 1.36$
& $-0.90 \pm 1.50$ \\
$\alpha$ 
& $0.98 \pm 0.09$ 
& $1.00 \pm 0.06$ 
& $0.96 \pm 0.21$ 
& $0.97 \pm 0.20$ 
& $1.01 \pm 0.02$
& $1.03 \pm 0.217$
& $0.99 \pm 0.26$ \\
\hline
\end{tabular}

    \caption{Mean best-fit of the model parameters, $\bar{\Theta}_{\text{model}}$, obtained sampling the \textsc{MD-Patchy}  and \textsc{Nseries} mocks of BOSS DR 12 galaxies, and \textsc{EZmocks} for eBOSS DR 16 QSO. For the \textsc{Nseries} mocks, the parameters correspond to an effective redshift $z_{\rm eff}=0.56$, which is slightly different from $z_3=0.61$, so the comparison is not exact.}
    \label{tab:BestFitParametersMocks}
\end{table}

 	\begin{table}[t]
 		\centering
\begin{tabular}{|c|c|c|c|}
\hline
& \textbf{Parameter} & \textbf{\textit{Model priors}} & \textbf{Value} \\
\hline\hline
\multirow{7}{*}{$\Theta_{\text{model}}$} 
& $b$              & Fixed             & 1 \\
& $\Sigma_{\text{s}}$   & Gaussian                & See \cref{tab:BestFitParametersMocks} \\
& $\Sigma_{\text{nl}}$  & Gaussian                & See \cref{tab:BestFitParametersMocks} \\
& $a_1$            & Gaussian                & See \cref{tab:BestFitParametersMocks} \\
& $a_2/10^3$       & Gaussian                & See \cref{tab:BestFitParametersMocks} \\
& $a_3/10^3$       & Gaussian                & See \cref{tab:BestFitParametersMocks} \\
& $\alpha$         & Uniform     & Best fit  $\pm 0.1$ (see \cref{tab:BestFitParametersMocks}) \\
\hline
\multirow{4}{*}{$\Theta_{\text{knots}}$}
& $y_1$            & Uniform     & $[-23,-17]$ \\
& $y_N$            & Uniform     & $[-23,-17]$ \\
& $x_i$            & Uniform     & $[-23,-17]$ \\
& $y_i, i \neq 1,N$            & Uniform     & $[-23,-17]$ \\
\hline
\multirow{3}{*}{$\Theta_{\text{cosmology}}$}
& $H_0$            & Fixed                   & 67.36 \\
& $\Omega_c h^2$   & Fixed                   & 0.1200 \\
& $\Omega_b h^2$   & Fixed                   & 0.02237 \\
\hline
\end{tabular}
    \caption{Prior sets used for the $P_{\mathcal{R}}(k)$ reconstructions. The multivariate Gaussian priors on $\Theta_{\text{model}}$ are correlated, following central values and standard deviations from the \textsc{MD-Patchy} and \textsc{EZmocks} best fits, listed in \cref{tab:BestFitParametersMocks}. The fixed values of $\Theta_{\text{cosmology}}$ are set by the Planck DR3 mean values.}
\label{tab:PriorsSpecific}
	\end{table}

\begin{figure}[t]
\centering 
\includegraphics[width=.55\textwidth]{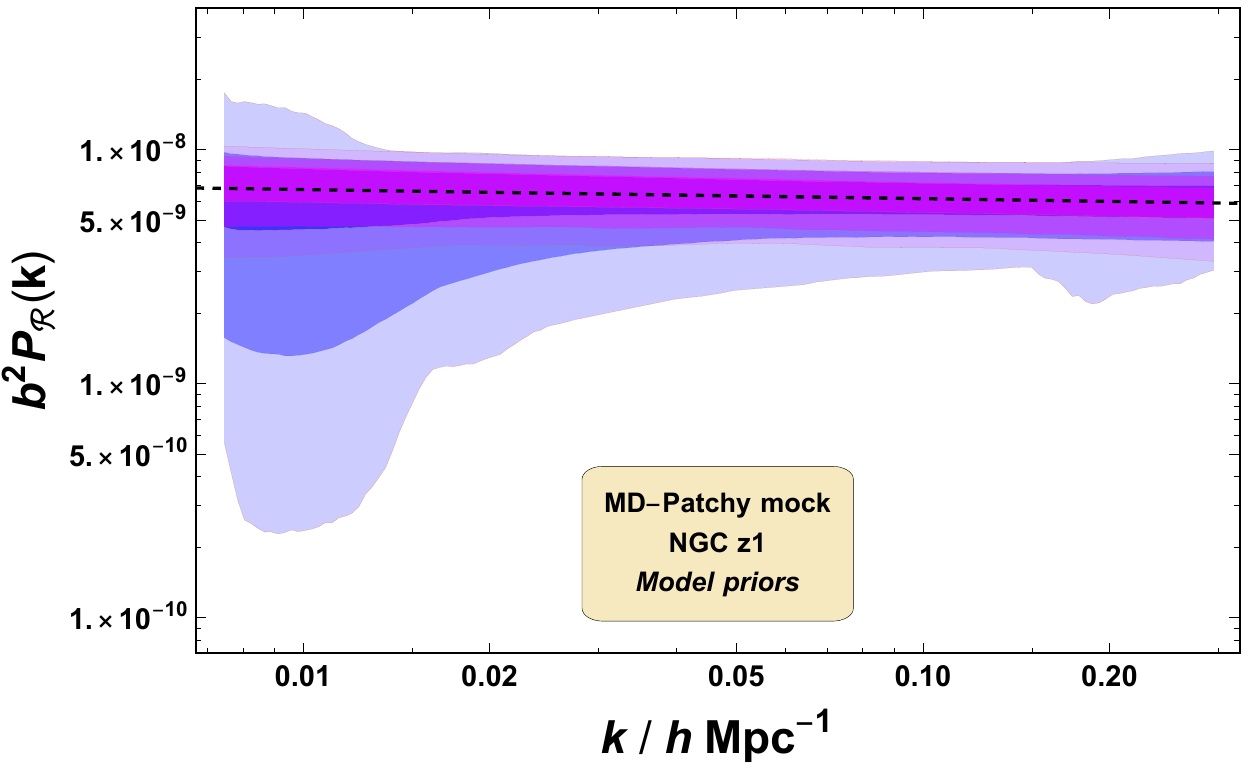}
\includegraphics[width=.30\textwidth]{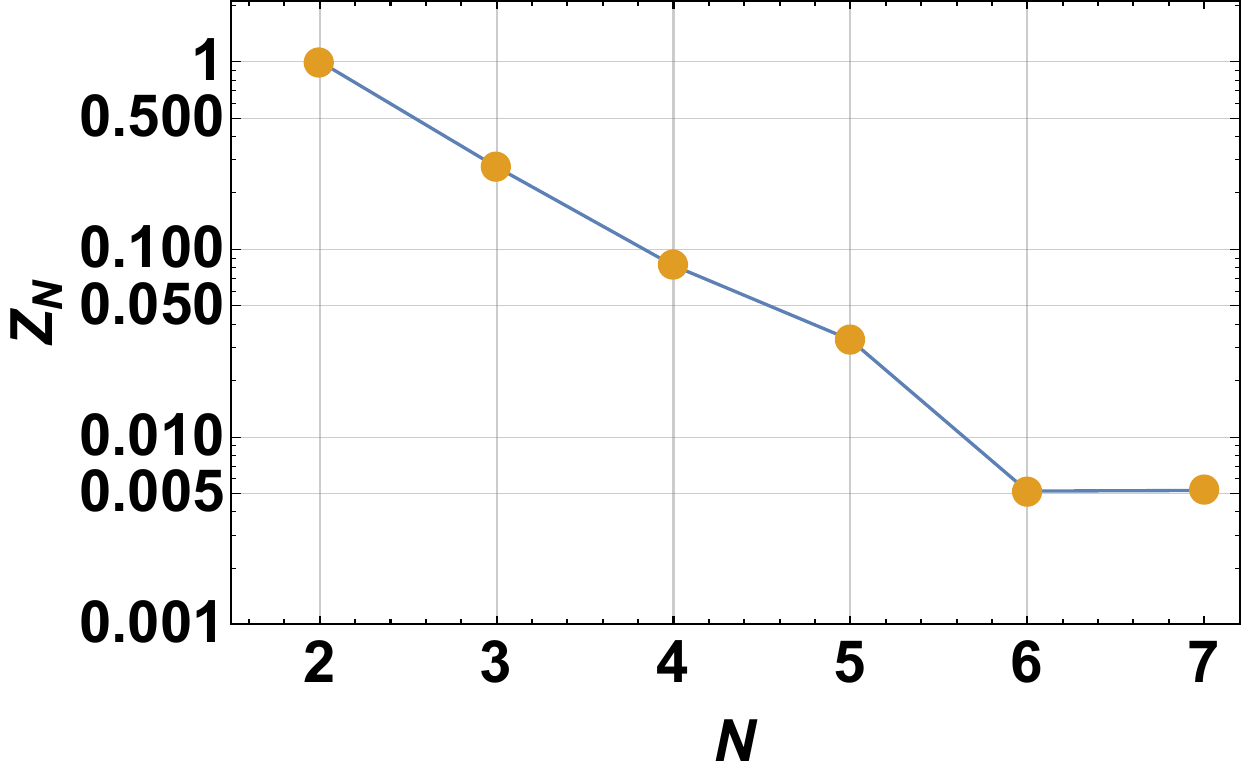}
\includegraphics[width=.55\textwidth]{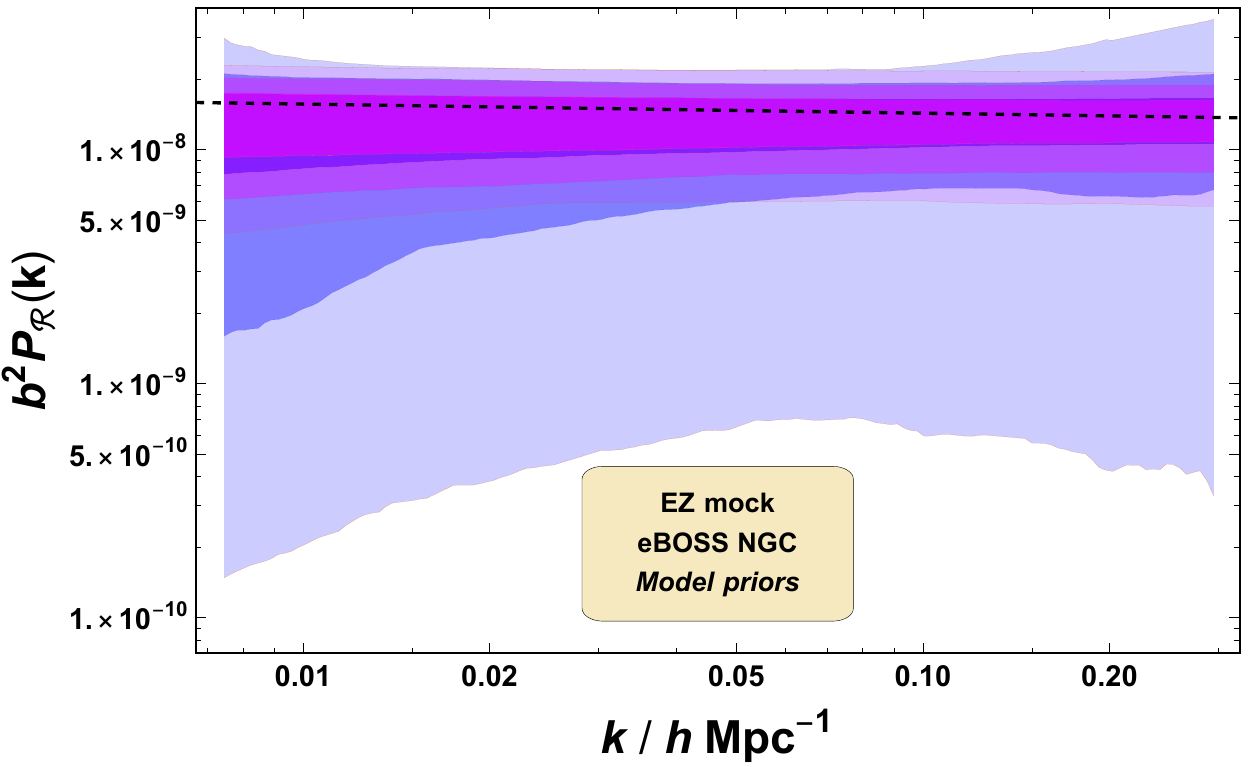}
\includegraphics[width=.30\textwidth]{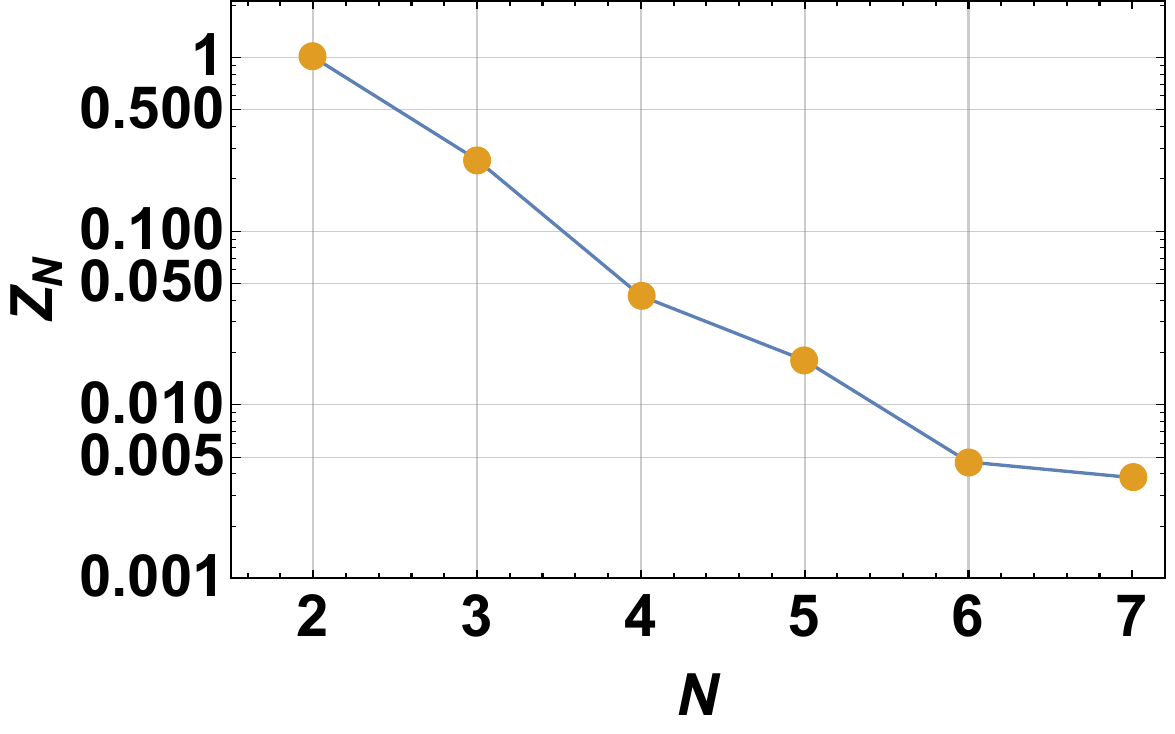}
\\[0.6em]
\caption{Primordial power spectrum reconstructions and their evidences applied to the \textsc{MD-Patchy} mock \#2001 of BOSS DR 12 NGC $z_1$ (top panels) and to the \textsc{EZmocks} \#891 of eBOSS DR 16 NGC (bottom panels). All results are obtained assuming \textit{model priors}. Note that the amplitude of $P_{\mathcal{R}}(k)$ is completely degenerate with $b^2$. The $N$-marginalized and $2$ knots contours are shown in blue and magenta, respectively. The dashed black line indicates the fiducial power-law of the mocks, corrected by $b^2$ for consistency.}
\label{fig:MocksNGCz1}
\end{figure}

We use some of the mocks not employed to infer the \textit{model priors} to reconstruct their primordial power spectra. In \cref{fig:MocksNGCz1} we show the reconstruction of $P_{\mathcal{R}}(k)$ for one \textsc{MD-Patchy} and one \textsc{EZmocks}. Their input power-law is successfully recovered for both mocks. The broadening of the contours, most prominent at large scales for the \textsc{MD-Patchy} mock, mainly arises from the 4-knot reconstruction, whose evidence remains high enough to still contribute to the $N$-marginalized posteriors. Similar reconstructions are obtained for different $z$-bins and also employing the \textsc{Nseries} mocks. In all cases the input power-law is recovered. 

\section{Results}
\label{sec:Results}

In this section, we apply the primordial power spectrum reconstruction framework to the BOSS and eBOSS data, using the \textit{model prior} set introduced in the validation. If we sample $\Theta_{\text{cosmology}}$ with Gaussian priors, their posteriors are dominated by the input priors. Therefore, we keep $\Theta_{\text{cosmology}}$ fixed during the reconstructions.

First, we perform the $P_{\mathcal{R}}(k)$ reconstruction for every individual redshift bin and galactic cap. Then, we perform a reconstruction jointly sampling the combined information from all the bins and caps, which is the most precise $P_{\mathcal{R}}(k)$ constraint in this work. The corresponding Bayes factors for all these explored scenarios are summarized in \cref{tab:BayesFactorPriors}. Finally, we infer the resulting $n_s$ values obtained under both prior choices, and compare them with the Planck DR3 values in \cref{fig:NsComparison}.
 	\begin{table}[t]
 		\centering
\begin{tabular}{|c|c|}
\hline
\textbf{Cap and $z$-bin} & \textbf{Bayes factor $Z_2/Z_{\text{max}}$} \\
\hline
BOSS NGC $z_1$ & $3.4$  \\
BOSS SGC $z_1$ & $4.3$  \\
BOSS NGC $z_3$ & $5.9$  \\
BOSS SGC $z_3$ & $7.9$  \\
\hline
Combined BOSS & $21.8$ \\
\hline
eBOSS NGC & $3.3$  \\
eBOSS SGC & $4.3$  \\
\hline
Combined eBOSS & $6.9$ \\
\hline
\textbf{Combined BOSS+eBOSS} & $26.1$ \\
\hline
\end{tabular}
 		\caption{Bayes factors of the $P_{\mathcal{R}}(k)$ reconstructions for each bin, and their combination. Except for the BOSS NGC $z_3$, the second biggest evidence is always $Z_3$.}
\label{tab:BayesFactorPriors}
	\end{table}

\subsection{Reconstruction results}
\label{subsec:ResultsReconstructions}

We start by reconstructing $P_{\mathcal{R}}(k)$ with the least prior information possible: using uniform priors for both $\Theta_{\text{model}}$ and $\Theta_{\text{knots}}$. We obtain a preference for a power-law, but with very poor precision in the reconstruction contours, as also obtained in the methodological validation.

\begin{figure}[t]
\centering 
\includegraphics[width=.30\textwidth]{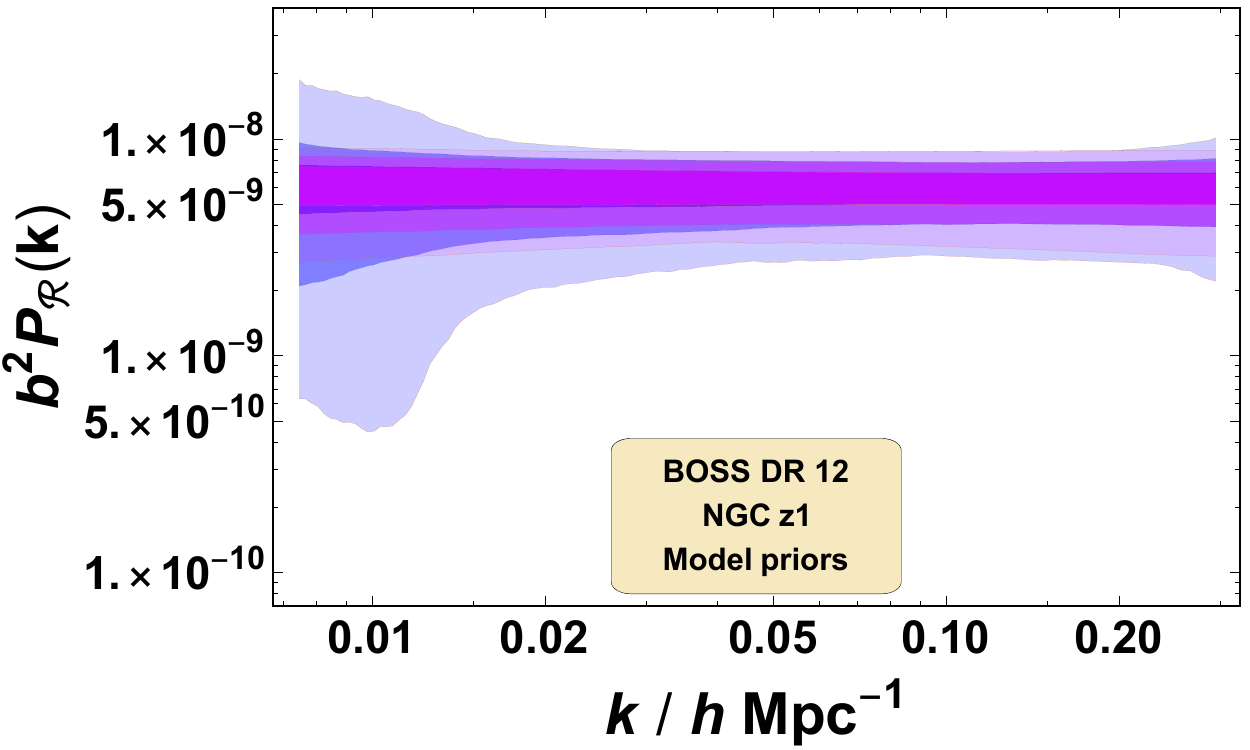}
\includegraphics[width=.18\textwidth]{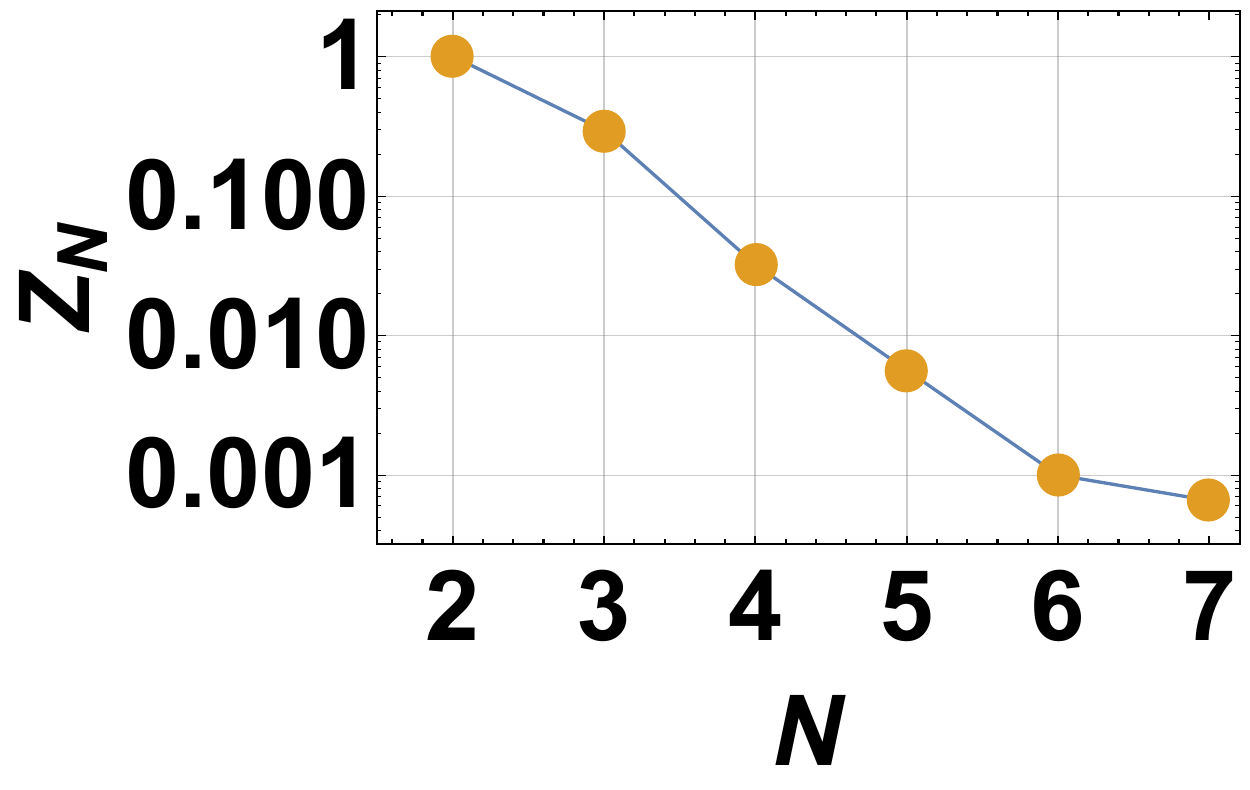}
\includegraphics[width=.30\textwidth]{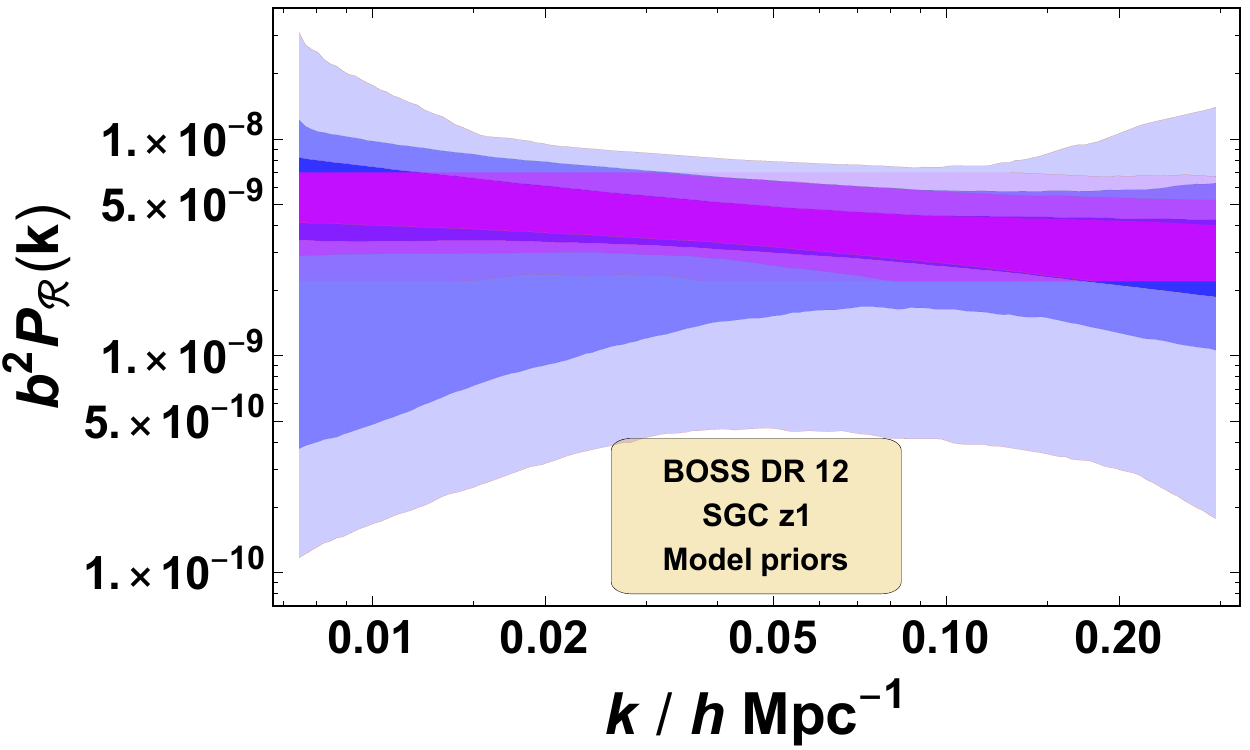}
\includegraphics[width=.18\textwidth]{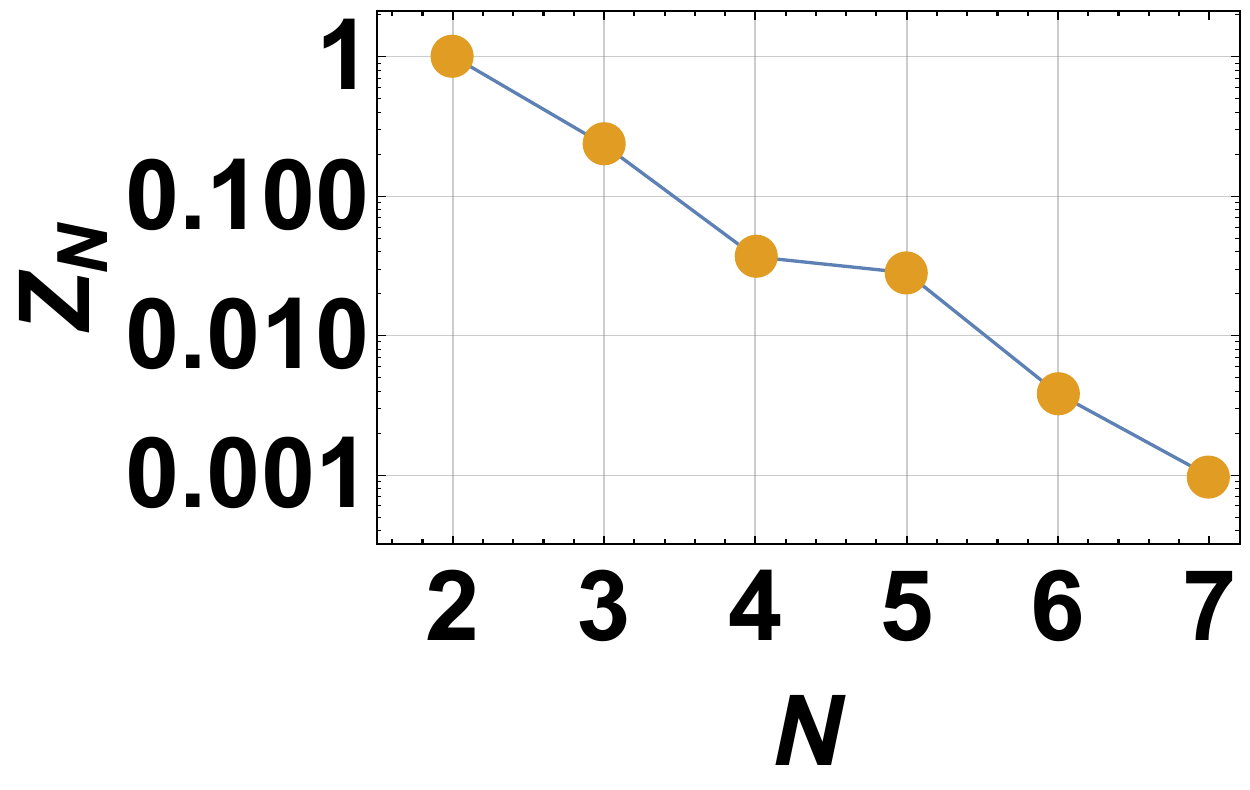}
\\[0.6em]
\includegraphics[width=.30\textwidth]{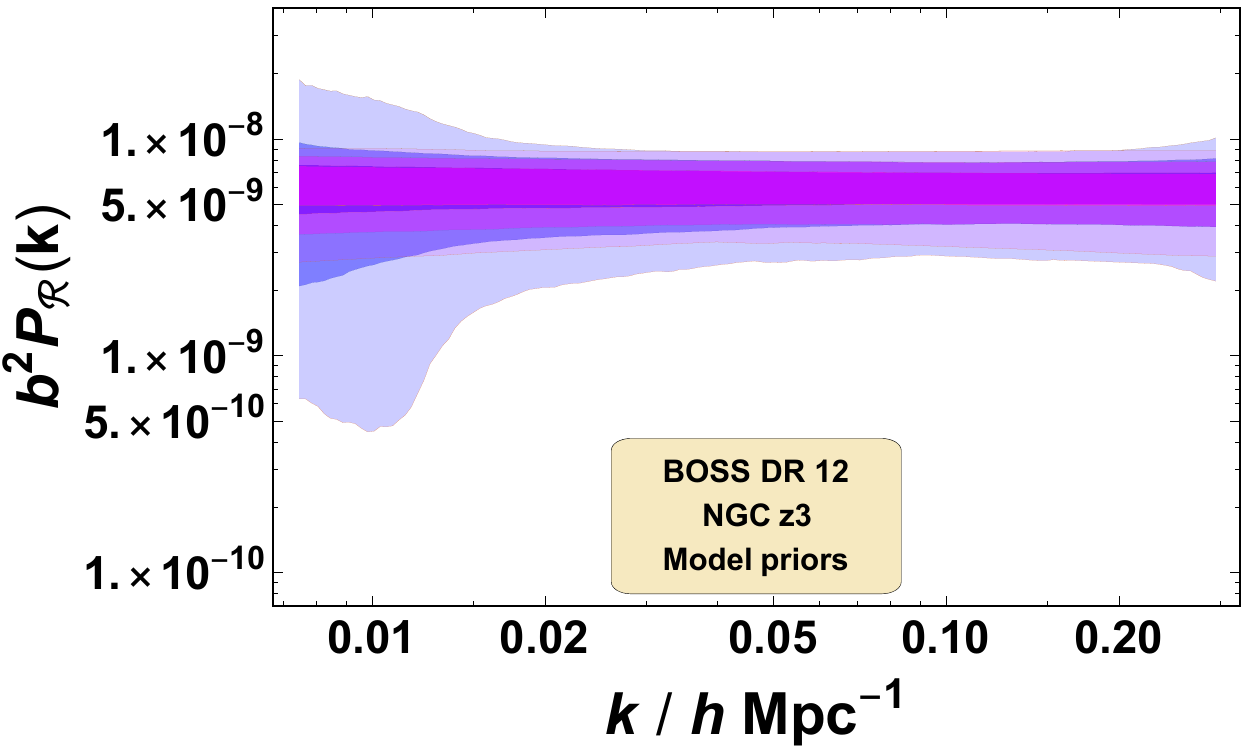}
\includegraphics[width=.18\textwidth]{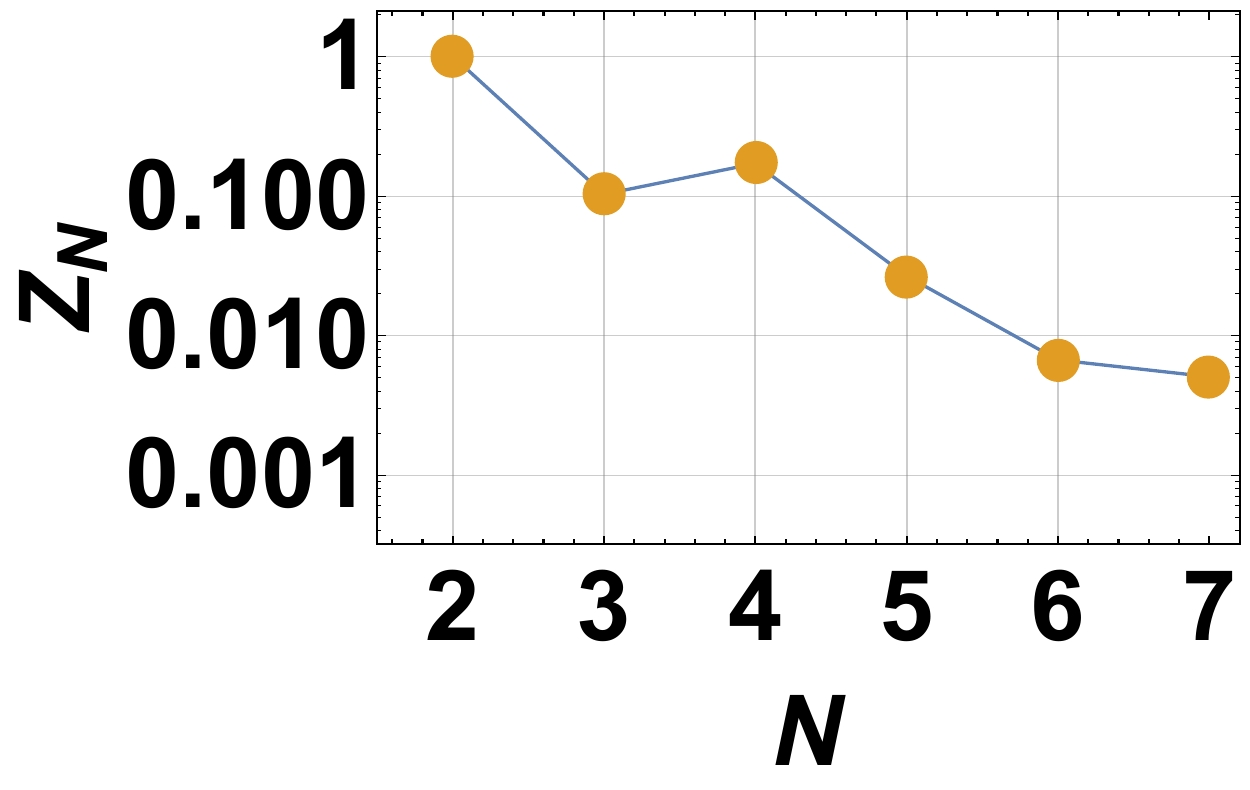}
\includegraphics[width=.30\textwidth]{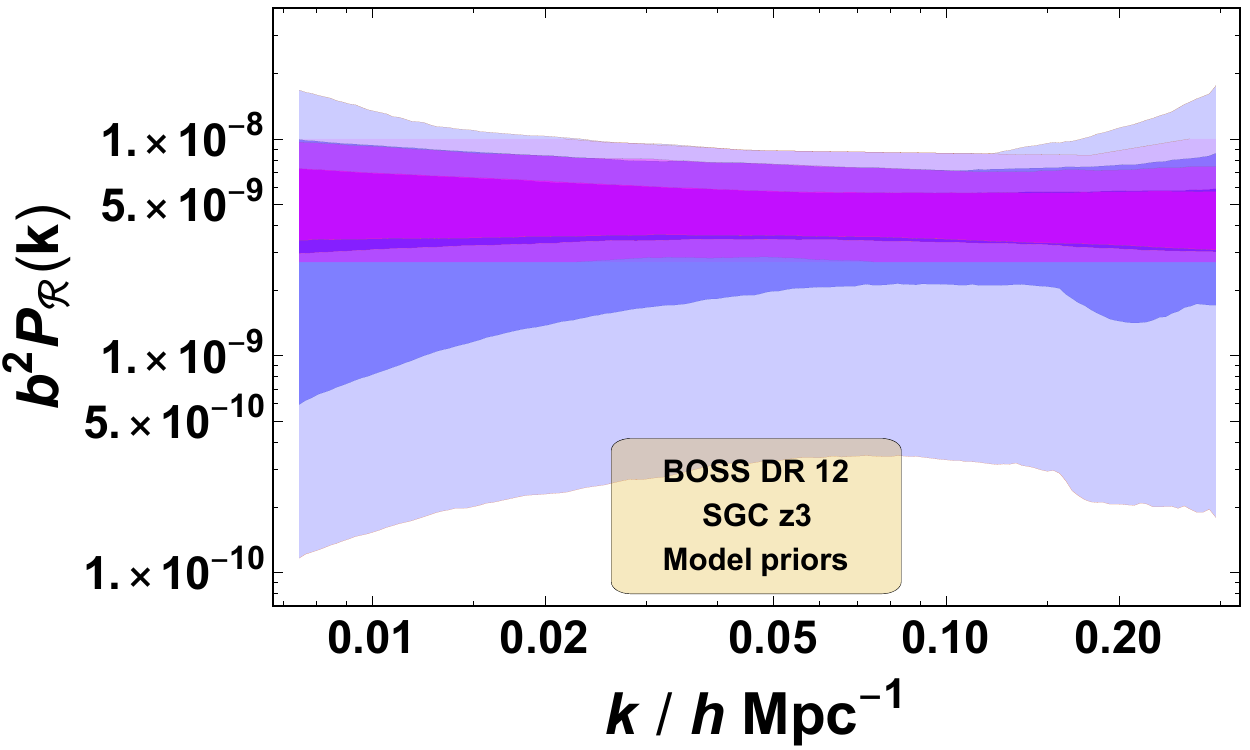}
\includegraphics[width=.18\textwidth]{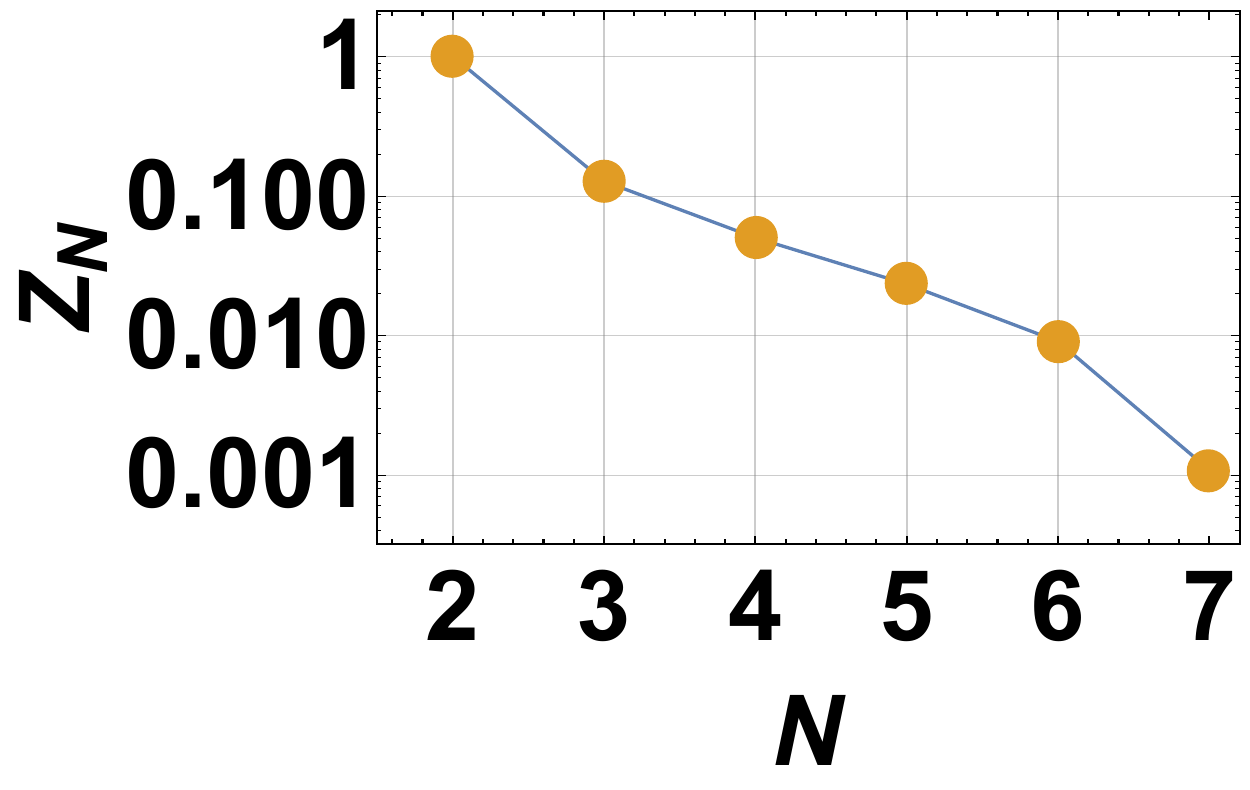}
\\[0.6em]
\includegraphics[width=.30\textwidth]{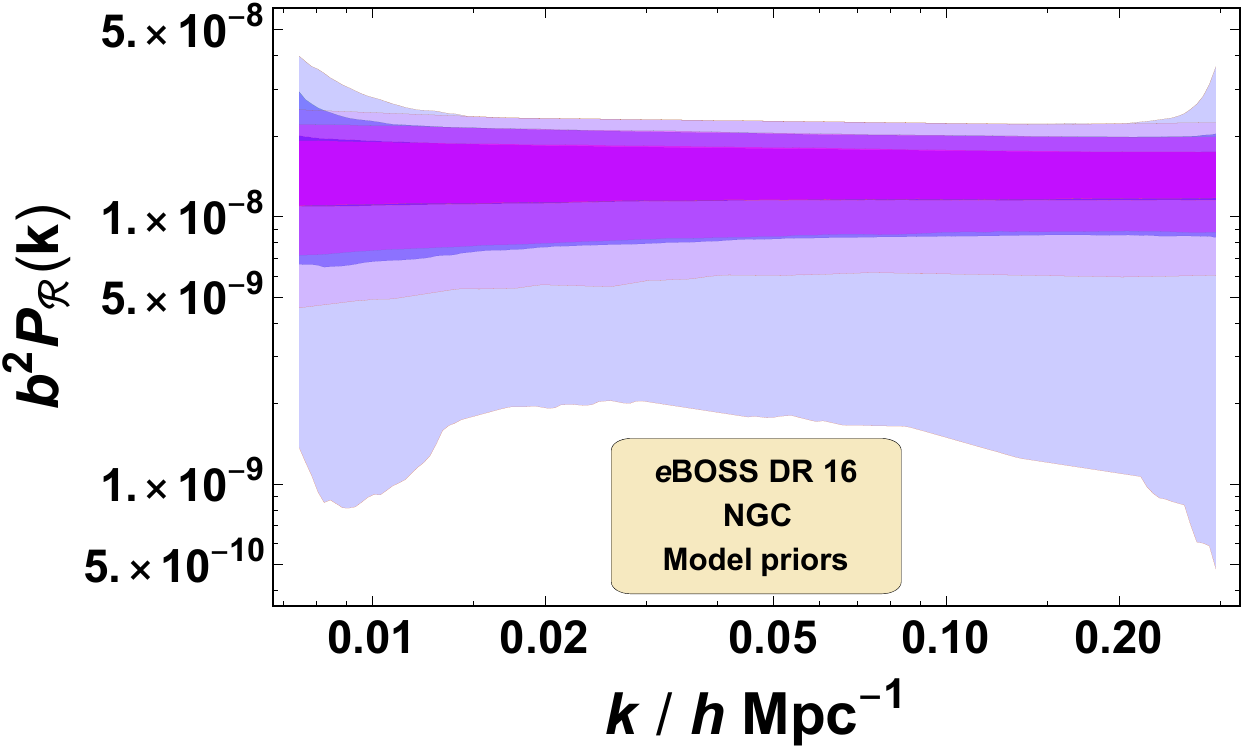}
\includegraphics[width=.18\textwidth]{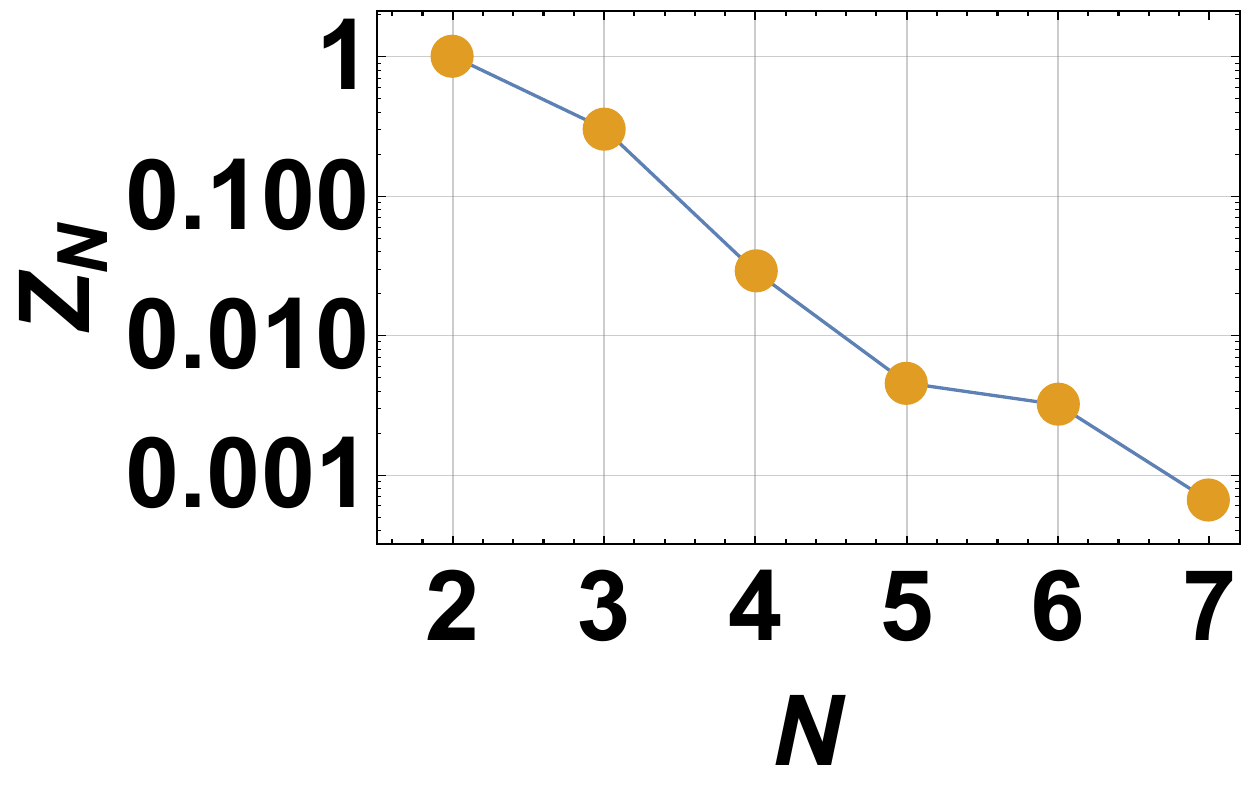}
\includegraphics[width=.30\textwidth]{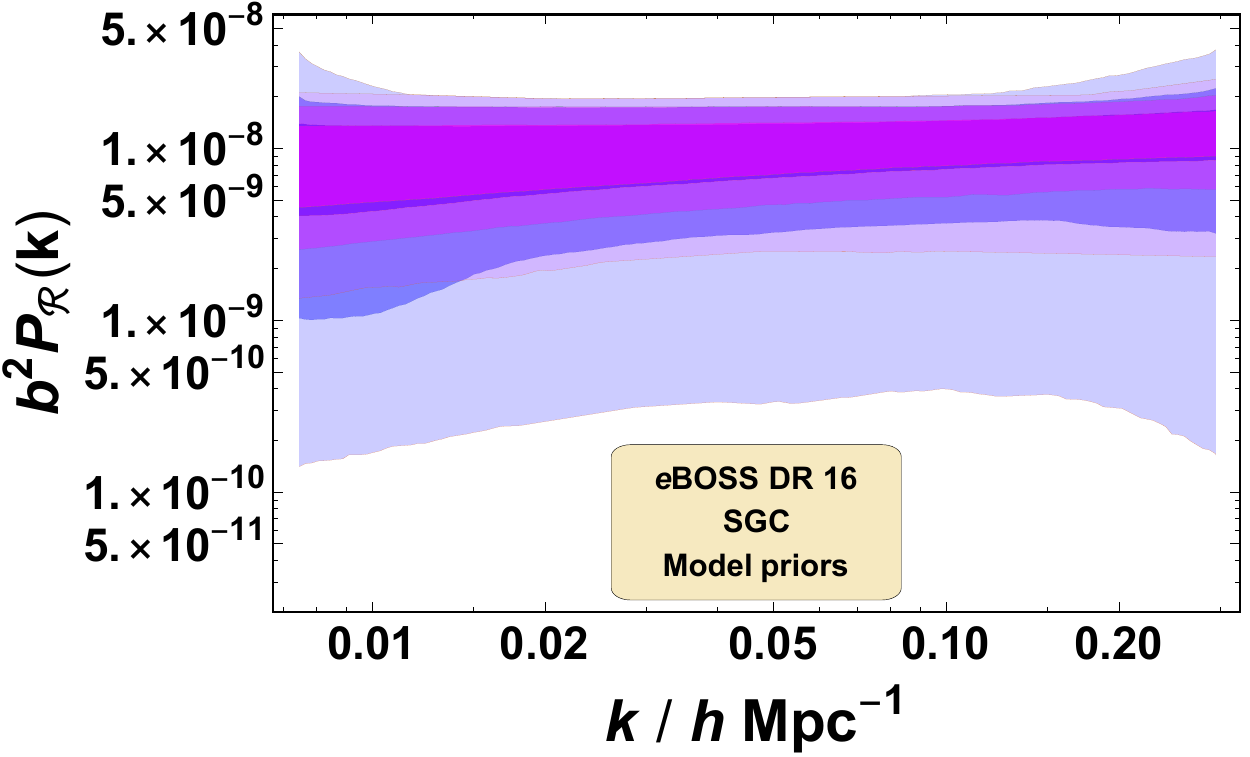}
\includegraphics[width=.18\textwidth]{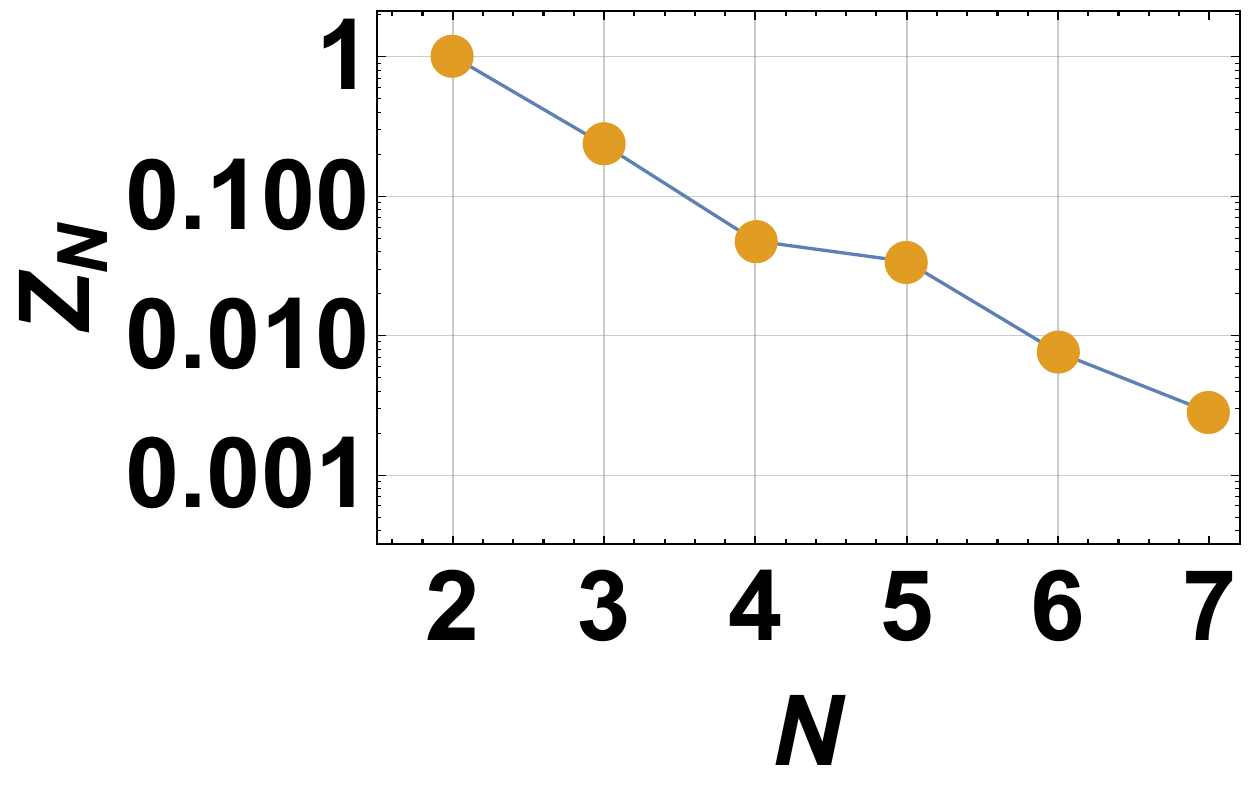}
\caption{Primordial power spectrum reconstructions for the BOSS DR 12 (top and middle panels) and eBOSS DR 16 galaxy bins (bottom panels). We assume the \textit{model priors} for all the scenarios. In magenta, the power-law contours, and in blue the $N$-marginalized contours. The evidences $Z_N$ for each case are also represented next to each reconstruction.}
\label{fig:SinglezbinModelPriors}
\end{figure}

To improve the precision and mitigate degeneracies between the knots and the model, we switch to the previously discussed \textit{model priors} (see \cref{tab:PriorsSpecific}). The obtained reconstructions are shown in \cref{fig:SinglezbinModelPriors} for each individual $z$-bin and cap. As in the validation, the amplitude of the spectrum is completely degenerate with $b^2$, so we show $b^2 P_{\mathcal{R}}(k)$ in the vertical axis. We find a similar performance to that obtained with the mocks: a power-law is recovered with statistical significance for all bins, but the $2\sigma$ and, especially, $3\sigma$ contours, remain broad. The most precise $P_{\mathcal{R}}(k)$ reconstruction is obtained for the NGC $z_1$ bin, while the SGC and eBOSS bins yield less precise contours. All the reconstructions show a clear and consistent preference for a primordial power-law spectrum.

\begin{figure}[t]
\centering 
\includegraphics[width=.72\textwidth]{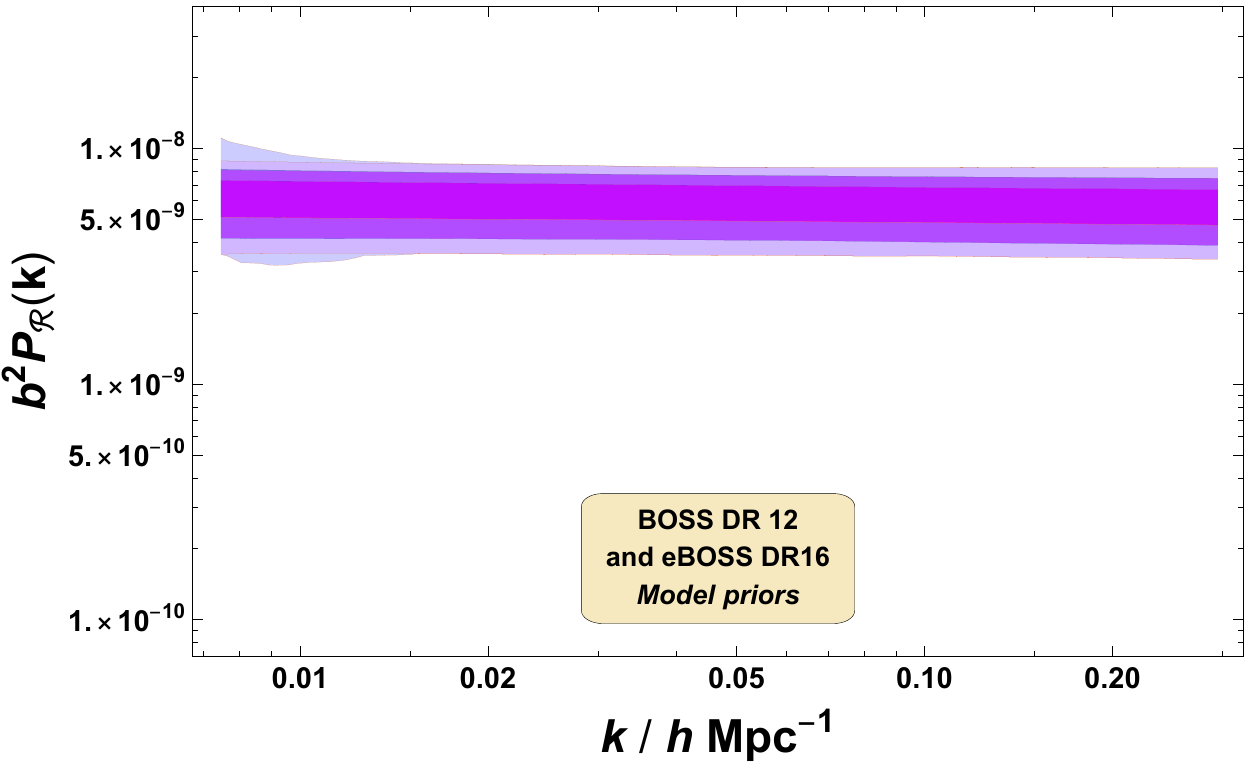}
\includegraphics[width=.41\textwidth]{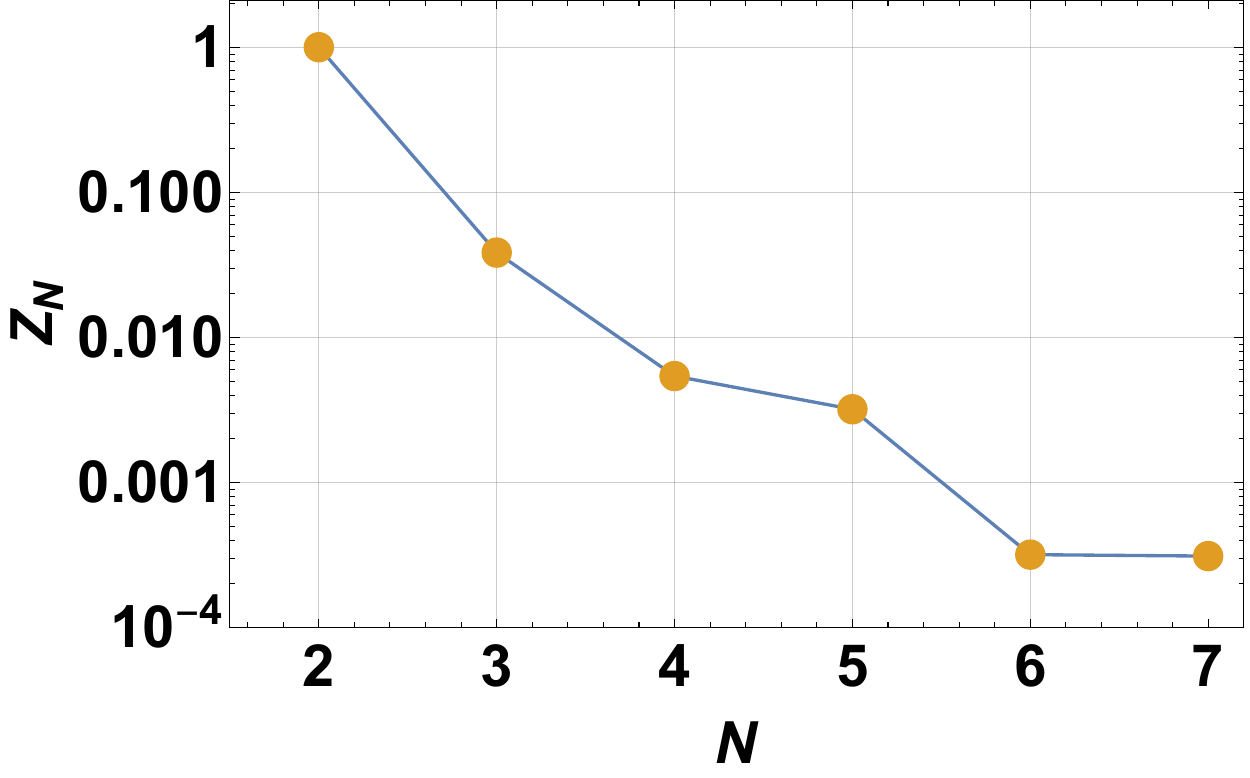}
\caption{Top: primordial power spectrum reconstruction for the combined BOSS DR 12 galaxies and eBOSS DR 16 QSO, assuming the \textit{model priors}. Both $N$-marginalized and $2$ knot contours, displayed in blue and magenta respectively, overlap almost completely. Bottom: evidence values for this combined scenario.}
\label{fig:CombinedModelPriors}
\end{figure}

In \cref{fig:CombinedModelPriors} we show the $P_{\mathcal{R}}(k)$ reconstruction obtained by combining all bins from BOSS and eBOSS. The $2\sigma$ and $3\sigma$ contours are much narrower, and the $N$-marginalized reconstruction is similar to the 2-knot one. The Bayes factor also indicates a strong preference for the 2-knot scenario, increasing by about a factor of 5 relative to the individual bins. All these results clearly point to the power-law case as the statistically most favored description of $P_{\mathcal{R}}(k)$.

\subsection{Constraints on $n_s$}
\label{subsec:ResultsNs}

Since no signs of PF have been detected, we constrain $n_s$ using the 2-knot $P_{\mathcal{R}}(k)$ reconstructions, which are equivalent to a power-law configuration. Then, we estimate the importance-weighted median values and quantiles from the sampled chains. The results are summarized in \cref{tab:nsAndAsConstraints} and represented in \cref{fig:NsComparison}. In the Appendix~\ref{subsec:RobustnessNs} we assess the robustness of the inferred $n_s$ values.
\begin{table}[t]
\renewcommand{\arraystretch}{1.15}
\centering
\begin{tabular}{|l|c|c|}
\hline
\textbf{Scenario} & \multicolumn{1}{|c|}{$n_s \pm 1\sigma \text{ } (3\sigma)$} & 
\multicolumn{1}{|c|}{$b^2 A_s / 10^{-9} \pm 1\sigma \text{ }(3\sigma)$} \\
\hline
BOSS+eBOSS & $0.976 \pm 0.021\ (^{+0.079}_{-0.063})$ & $6.0^{+0.9}_{-1.1}\ (^{+2.3}_{-2.5})$ \\
BOSS & $0.972 \pm 0.028\ (^{+0.11}_{-0.09})$ & $5.9^{+0.9}_{-1.0}\ (^{+2.3}_{-2.5})$ \\
Planck+BOSS+eBOSS & $0.9652 \pm 0.0041$ & -- \\
\hline
Planck DR3 & $0.9649 \pm 0.0042$ & $2.099 \pm 0.030$ \\
\hline
\end{tabular}
\caption{Constraints on the spectral index $n_s$ and amplitude $A_s$ (or $b^2A_s$) from the 2-knot $P_{\mathcal{R}}(k)$ reconstructions. All inferred values are consistent with Planck DR3 TT, TE, EE+lowE+lensing, shown in the last row.}
\label{tab:nsAndAsConstraints}
\end{table}
\begin{figure}[t]
\centering 
\includegraphics[width=.79\textwidth]{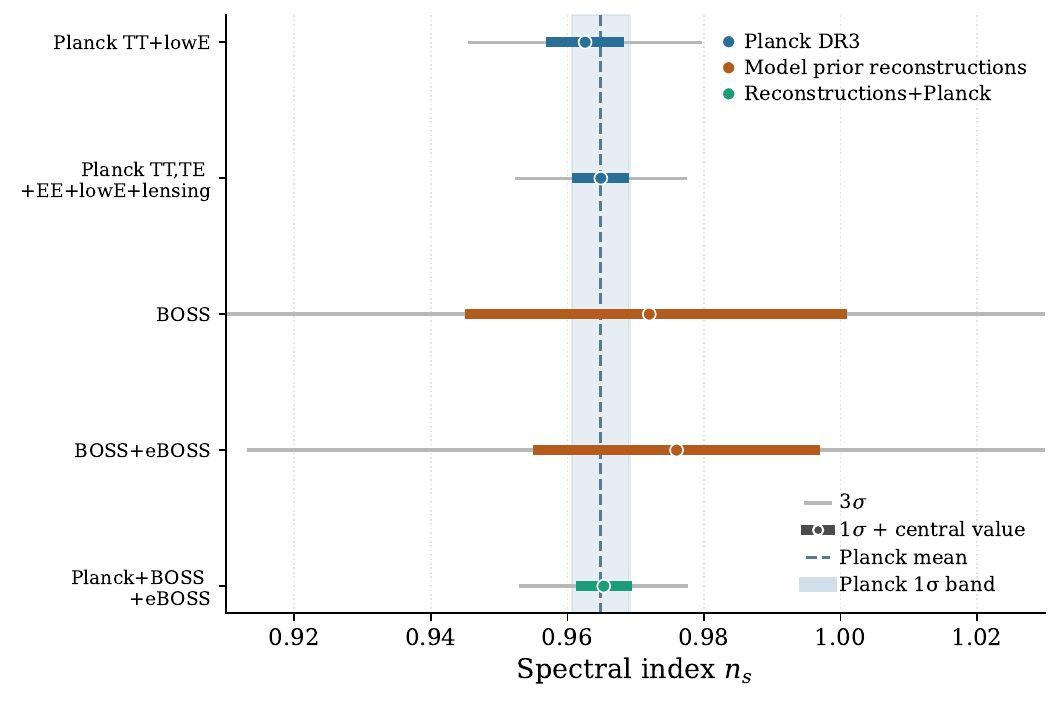}
\caption{Constraints on $n_s$, together with the corresponding Planck DR3 reference values, and their independent joint combination. Thick colored bars denote the $1\sigma$ intervals, thin lighter bars the $3\sigma$ intervals, and the markers indicate the central values. The vertical dashed line and shaded band show the mean and $1\sigma$ region of the Planck DR3 TT,TE,EE+lowE+lensing constraint. The different colors label the various dataset combinations.}
\label{fig:NsComparison}
\end{figure}

All the obtained constraints are compatible with Planck DR3 even at the $1\sigma$ level. In particular, the obtained value for $n_s$ from the BOSS+eBOSS $P_{\mathcal{R}}(k)$ reconstructions is:
\begin{equation}
\label{NsBOSSandeBOSS}
n_s = 0.976 \pm 0.021\,(1\sigma),
\end{equation}
which is compatible with the Planck DR3 TT,TE,EE+lowE+lensing value of $n_s=0.9649\pm0.0042$. Combining this Planck value with the BOSS+eBOSS constraint using an inverse‑variance weighted average yields:
\begin{equation}
\label{NsCombinedPlanck}
n_s = 0.9652 \pm 0.0041\,(1\sigma),
\end{equation}
with a central value shift of $+0.0003$ and $\sim2\%$ error reduction compared to Planck alone.

\section{Conclusions}
\label{sec:Conclusions}

The presence of Primordial Features (PF) in the primordial power spectrum $P_{\mathcal{R}}(k)$ is a powerful probe of the inflationary epoch, since many models departing from slow-roll predict a variety of such features. In this work, we have reconstructed the $P_{\mathcal{R}}(k)$ of the BOSS DR 12 LRG and eBOSS DR 16 QSO catalogs with a non-parametric Bayesian knot technique, replacing the Standard Model power law with a set of knot parameters $\Theta_{\text{knots}}$.

In contrast to previous applications, we apply the method to a parametric model of the galaxy power spectrum. This model assumes any power-law deviations and BAO contributions separately from the matter power spectrum template, following standard BAO analyses. The parametrization encompasses a bias, a set of three broadband parameters, and additional RSD and BAO effects, with a total of 7 parameters $\Theta_{\text{model}} = \{b, a_1, a_2, a_3, \Sigma_s, \Sigma_{\text{nl}}, \alpha\}$. In this way, we are able to reconstruct $P_{\mathcal{R}}(k)$ at non-linear scales, up to $0.3 \text{ h} \text{ Mpc}^{-1}$.

We first assess the sensitivity of the method to PF with the parametric description: local oscillatory features can be identified for amplitudes as low as 5\%. We then apply the $P_{\mathcal{R}}(k)$ reconstructions to \textsc{MD-Patchy} and \textsc{EZmock} mock catalogs, representative of the BOSS LRG and eBOSS QSO catalogs, respectively. In all cases, we successfully recover their input power law $P_{\mathcal{R}}(k)$ for different redshift bins and mock realizations. Based on these mock analyses, we also derive a set of correlated Gaussian priors from their mean best fits, named \textit{model priors}, which help reduce degeneracies between $\Theta_{\text{knots}}$ and $\Theta_{\text{model}}$ and improve the reconstruction precision. For an even more robust validation, mocks generated with different PF templates would be a valuable extension of this work.

The results for the individual $z$-bins and galactic caps of the BOSS DR 12 LRG and eBOSS DR 16 QSO are consistent, showing no evidence of the presence of PF in $P_{\mathcal{R}}(k)$ and pointing to the power law as the preferred model for $P_{\mathcal{R}}(k)$. Combining all these independent redshift bins and galactic caps enhances the reconstruction precision, yielding a preference for the power law with stronger statistical evidence. Additionally, we test the robustness of this conclusion by reconstructing $P_{\mathcal{R}}(k)$ using an alternative prior set, in which most Gaussian constraints on $\Theta_{\text{model}}$ are relaxed and Planck-based priors are imposed on the extreme knots. This analysis also favors the power law with statistical significance. 

Since no evidence for PF is found, we robustly estimate the spectral index from the reconstructions, obtaining $n_s = 0.976 \pm 0.021$, which is compatible with the Planck result, albeit with an uncertainty five times larger. The amplitude $A_s$ cannot be independently constrained, as it is fully degenerate with the bias $b$. The obtained value of $n_s$ points to a quasi-scale invariant power law for $P_{\mathcal{R}}(k)$, as predicted by most slow-roll inflationary models.

In this work we explore scales $k \in [0.007, 0.3] \text{ h} \text{ Mpc}^{-1}$. The BOSS and eBOSS data are rebinned from 400 to 40 $k$-modes to facilitate the complex sampling required to account for selection function effects and inter-scale correlations. Using the full set of scales, this range could be extended down to $k \in [0.0018, 0.3] \text{ h} \text{ Mpc}^{-1}$, allowing PF to be probed over a larger number of modes. Combining independent redshift bins significantly enhances the $P_{\mathcal{R}}(k)$ reconstruction, and the inclusion of additional bins and/or tracers could further improve the results, provided correlations between non-independent samples are properly modeled. Looking ahead, Stage-IV surveys such as DESI or Euclid are expected to deliver constraints on $P_{\mathcal{R}}(k)$ that surpass those of Planck.

\acknowledgments 
The authors acknowledge Santander Supercomputacion support group at the University of Cantabria who provided access to the supercomputer Altamira Supercomputer at the Institute of Physics of Cantabria (IFCA-CSIC), member of the Spanish Supercomputing Network, for performing simulations/analyses. The authros also thank Red de Investigación RED2022-134715-T funded by 
MCIN/AEI/10.13039/501100011033. GMS, AMC, and EMG also thank the Spanish AEI and MICIU for the financial support provided under the projects with references PID2019-110610RB and PID2022-139223OB-C21, and acknowledge support from Universidad de Cantabria and Consejería de Universidades, Igualdad, Cultura y Deporte del Gobierno de Cantabria via the \textit{Instrumentación y ciencia de datos para sondear la naturaleza del universo} project. Those authors also acknowledge financial support from the Complementary Plan in Astrophysics and High-Energy Physics (CA25944), project C17.I02.P02.S01.S03 CSIC, supported by the Next Generation EU funds, RRF and PRTR mechanisms, and the Government of the Autonomous Community of Cantabria. GMS acknowledges financial support from the Formación de Personal Investigador (FPI) programme, ref. PRE2018-085523, associated to the Spanish Agencia Estatal de Investigación (AEI, MICIU) project ESP2017-83921-C2-1-R, and also from the project UC-LIME (PID2022-140670NA-I00), financed by MCIN/AEI/ 10.13039/501100011033/FEDER, UE. HGM acknowledges support through the Consolidación Investigadora (CNS2023-144605) of the Spanish Ministry of Science and Innovation. We acknowledge the use of PolyChord \cite{MandatoryPolyChord1,MandatoryPolyChord2}, CAMB \cite{CAMB}, GetDist \cite{GetDist}, and Cobaya \cite{CobayaMandatory1,CobayaMandatory2}. The Cobaya code has been created with the help of the Python packages \texttt{NumPy} \cite{Numpy}, \texttt{Matplotlib} \cite{Matplotlib} and \texttt{SciPy} \cite{Scipy}.

\appendix

\section{BAO Smoothing}
\label{subsec:AppendixBAO}

Nonlinear matter clustering smooths out BAO in the matter power spectrum. In order to take into account this effect, a no-wiggle matter power spectrum $P_{nw}(k)$ is needed. 

To ensure robustness for cosmologies that deviate significantly from the fiducial model, such as those encountered during our sampling, we require a no-wiggle matter power spectrum prescription that remains reliable across a broad parameter space. The Savitzky–Golay filtering technique employed in \cite{SavGolExample} is not well suited to this task. Instead, we adopt the method introduced in \cite{BriedenNoWiggle}. This method identifies all local extrema of the gradient of the rescaled power spectrum $P_m(k)\allowbreak \text{ }k^{7/4}$, constructs separate interpolation curves through the maxima and minima, and defines the no-wiggle spectrum $P_{nw}(k,z)$ as their mean.

\section{Robustness of the $n_s$ constraints}
\label{subsec:RobustnessNs}

In order to assess the robustness of the obtained $n_s$ constraints, we performed a Fisher analysis based on the derivatives of the convolved galaxy power spectrum with respect to $n_s$ and the $\Theta_{\text{model}}$ parameters, evaluated at each $z$-bin for the fiducial best-fit model. For a galaxy power spectrum model vector, $\mathbf{P}$, and parameters $\theta_i$, we construct the Fisher matrix as:
\begin{equation}\label{FisherMatrix}
F_{ij} = \frac{\partial \mathbf{P}^{\mathrm{T}}}{\partial \theta_i}\,\mathbf{C}^{-1}\,\frac{\partial \mathbf{P}}{\partial \theta_j},
\end{equation}
where $\mathbf{C}$ is the covariance matrix of the data for each cap and $z$-bin. The analytical bias induced on the model parameters by a shift $\delta n_s$ is then estimated from the inverse Fisher matrix as:
\begin{equation}\label{LinearShift}
\Delta \theta_i \simeq \frac{(F^{-1})_{i n_s}}{(F^{-1})_{n_s n_s}}\,\delta n_s,
\end{equation}
We evaluate $\Delta \theta_i$ for each $z$-bin and cap, and compare them to their respective prior widths $\sigma_{\theta_i}$ (see \cref{tab:PriorsSpecific}). We explore a shift $\delta n_s = 0.042$, which encompasses $\pm 1\sigma$ deviations from our BOSS+eBOSS results (see \cref{tab:nsAndAsConstraints}), obtaining $\Delta \theta_i / \sigma_{\theta_i} < 0.26$ for all the $\Theta_{\text{model}}$ parameters, except $\Sigma_s$ and $a_3$. In the case of the eBOSS, this ratio exceeds 1 for these two parameters, and for the BOSS we find $\Delta \Sigma_s / \sigma_{\Sigma_s} < 0.07$ and $\Delta a_3 / \sigma_{a_3} < 0.9$ for all bins. Thus, we focus on the impact of the $a_3$ and $\Sigma_s$ priors on the inferred value of $n_s$.

Since widening the $a_3$ and $\Sigma_s$ priors affects the other $\Theta_{\text{model}}$ parameters (including their correlations), we investigate their impact. We start by diagonalizing each covariance matrix as $C = V D V^T$, where $D = \mathrm{diag}(y_1,\dots,y_5)$ contains the eigenvalues, and the columns of $V$ define the orthonormal modes $y_k$, sorted depending on their values. The $y_5$ and $y_4$ modes are the largest ones, mainly depending on $a_3$ and $a_2$, while $y_2$ is the mode with the largest $\Sigma_s$ contribution. Given the large correlations between $a_3$ and $a_2$ and the large value of the $y_4$ eigenvalue, the $a_2$ impact also has to be assessed to infer $n_s$. In summary, we want to explore the impact of $n_s$ that a broadening of the parameters $\{\Sigma_s, a_2, a_3\}$ induces, having identified them with the modes $\{y_4,y_5\}$ for BOSS and $\{y_2, y_4, y_5\}$ for eBOSS. We marginalize over these $y_i$ by constructing diagonal matrices $D_{\mathrm{marginalized}}$, constructed only with the modes $\{y_1,y_2,y_3\}$ for the BOSS galaxies, and with only $\{y_1,y_3\}$ for the eBOSS QSO. By keeping the remaining modes, we are able to preserve the $\Theta_{\text{model}}$ correlations on the rest of parameters. As the last step, we map these $D_{\mathrm{marginalized}}$ back to the original parameter space, via $C^{-1}_{\mathrm{marginalized}} = V D^{-1}_{\mathrm{marginalized}} V^T$. By this, we obtain an inverse covariance matrix $C^{-1}_{\mathrm{marginalized}}$ consistent with a prior set marginalized over $y_4$, $y_5$ and/or $y_2$, while preserving the correlation structure of the original covariance. We denote this wider set of priors as \textit{wider model priors}.

\begin{figure}[t]
\centering 
\includegraphics[width=.69\textwidth]{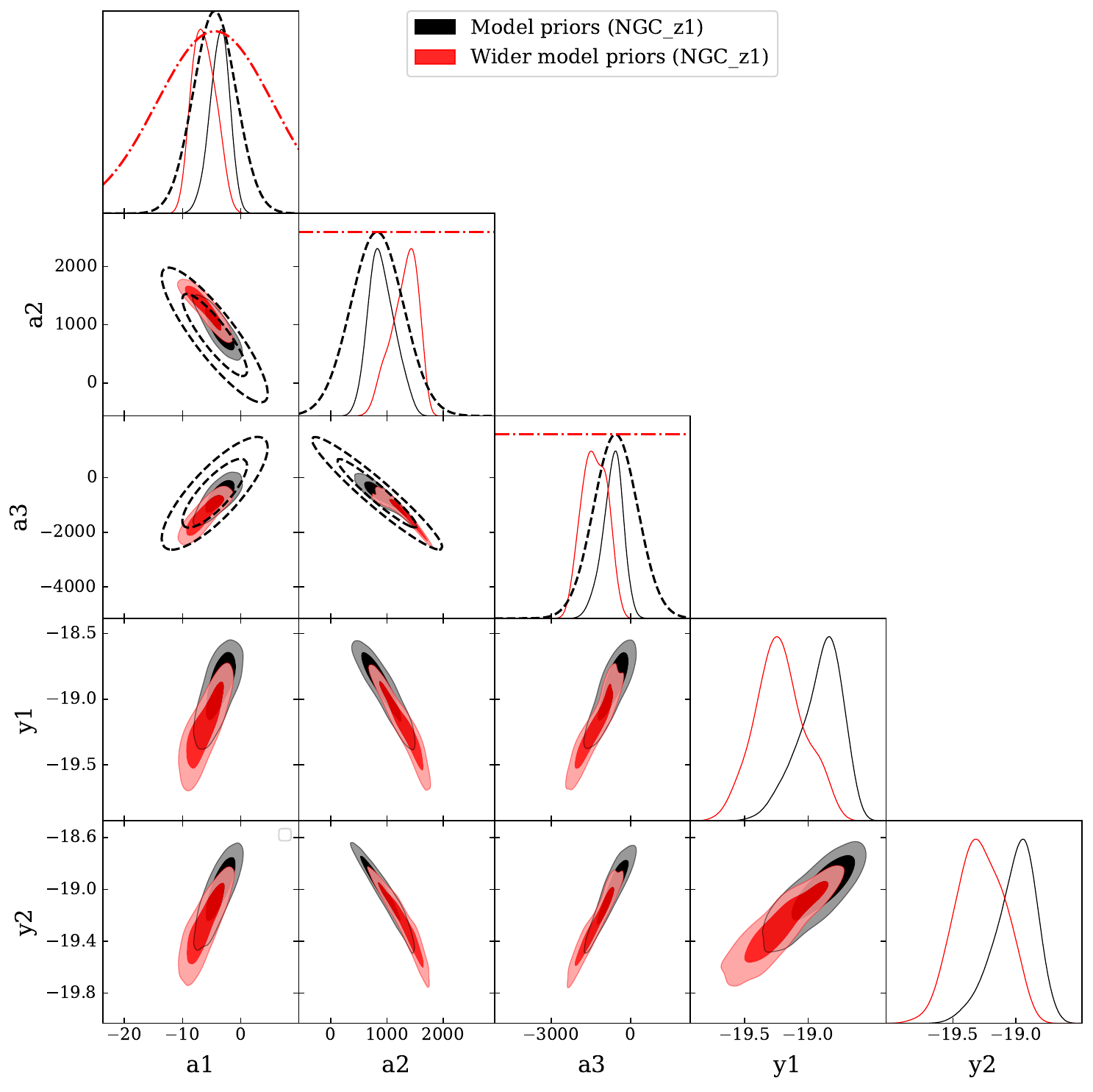}
\caption{Solid lines: posterior distributions for the model parameters $\{a_1,a_2,a_3\}$ and the knot parameters $\{y_1, y_2\}$. Dashed lines: prior distributions. In black we show the original \textit{model prior} set, and in red the \textit{wider model prior} set. The priors for the \textit{wider model priors} on $a_2$ and $a_3$ are not uniform, but they have been substantially broadened, causing them to appear nearly flat in the scales displayed in this plot.}
\label{fig:PosteriorsModelPriorsVsWiderModelPriors}
\end{figure}

We show in \cref{fig:PosteriorsModelPriorsVsWiderModelPriors} the priors of the broadband parameters $a_i$, and the posteriors of these parameters and the knot parameters $y_1$ and $y_2$. In the plot, the correlations among them are also displayed in the triangular figures. All posteriors and their correlations remain well inside the wider priors, without significant broadening. This points to a small effect on the reconstructions when the $a_i$ priors are broadened. This test also highlighted the importance of the correlations among $\Theta_{\text{model}}$.

In \cref{tab:nsRobustness} we constrain $n_s$ with the \textit{wider model priors}, obtaining a slight increase of the central value of $n_s$ and in its $1\sigma$ level. These small variations suggest that our choice of priors has only a minor impact on the determination of $n_s$. 
\begin{table}[t]
\renewcommand{\arraystretch}{1.15}
\centering
\begin{tabular}{|l|c|}
\hline
BOSS+eBOSS & \multicolumn{1}{|c|}{$n_s \pm 1\sigma \text{ } (3\sigma)$} \\
\hline
\textbf{\textit{Model priors}} & $0.976 \pm 0.021\ (^{+0.079}_{-0.063})$ \\
\textbf{\textit{Wider model priors}} & $0.983 \pm 0.026\ (^{+0.093}_{-0.098})$ \\
\hline
\end{tabular}
\caption{Constraints on the spectral index $n_s$ for the wider model prior set. The standard deviation of $n_s$ remains similar.}
\label{tab:nsRobustness}
\end{table}

\section{\textit{Power-law priors} reconstructions }
\label{subsec:PLPriors}

We explore another complementary prior set in order to doble check the obtained results and to improve the reconstruction precision. This prior set assumes informative constraints of the outermost knot amplitudes according to Planck DR3, while being kept minimally informative in $\Theta_{\text{model}}$, except for $\Sigma_{\text{nl}}$. We refer to it as \textit{power-law priors}. In \cref{tab:PriorsTwoSets} we provide the numerical values used for this prior set.
 	\begin{table}[t]
 		\centering
\begin{tabular}{|c|c|c|c|}
\hline
& \textbf{Parameter}  & \textbf{\textit{Power-law priors}} & \textbf{Values} \\
\hline\hline
\multirow{7}{*}{$\Theta_{\text{model}}$} 
& $b$                         & Uniform $[\pm 5\sigma]$      &  \\
& $\Sigma_{\text{s}}$                & Uniform $[\pm 3\sigma]$      &  \\
& $\Sigma_{\text{nl}}$                 & Gaussian      &  \\
& $a_1$                & Uniform $[\pm 3\sigma]$      & See \cref{tab:BestFitParametersMocks} \\
& $a_2/10^3$           & Uniform $[\pm 3\sigma]$      &  \\
& $a_3/10^3$           & Uniform $[\pm 3\sigma]$      &  \\
& $\alpha$              & Uniform $[\pm 0.1]$          &  \\
\hline
\multirow{4}{*}{$\Theta_{\text{knots}}$}
& $y_1$                & Gaussian                     & $-19.901 \pm 0.015$ \\
& $y_N$                & Gaussian                     & $-20.030 \pm 0.016$ \\
& $x_i$                & Uniform         & $[0,1]$ \\
& $y_i, i \neq 1,N$                & Uniform          & $[-23,-17]$ \\
\hline
\multirow{3}{*}{$\Theta_{\text{cosmology}}$}
& $H_0$                          & Fixed                    & 67.36  \\
& $\Omega_c h^2$                 & Fixed                    & 0.1200  \\
& $\Omega_b h^2$                 & Fixed                    & 0.02237 \\
\hline
\end{tabular}
    \caption{Priors for the reconstructions assuming the \textit{power-law} priors for $\Theta_{\text{model}}$. This prior set assumes informative constraints of the outermost knot amplitudes $\{y_1, y_N\}$ according to Planck DR3, while $\Theta_{\text{model}}$ is kept as non-informative as possible. The uniform priors on $\Theta_{\text{model}}$ indicate the central value with $3\sigma$ or $5\sigma$ deviations of the mocks best fit, listed in \cref{tab:BestFitParametersMocks}. }
\label{tab:PriorsTwoSets}
	\end{table}

\begin{figure}[t]
\centering 
\includegraphics[width=.50\textwidth]{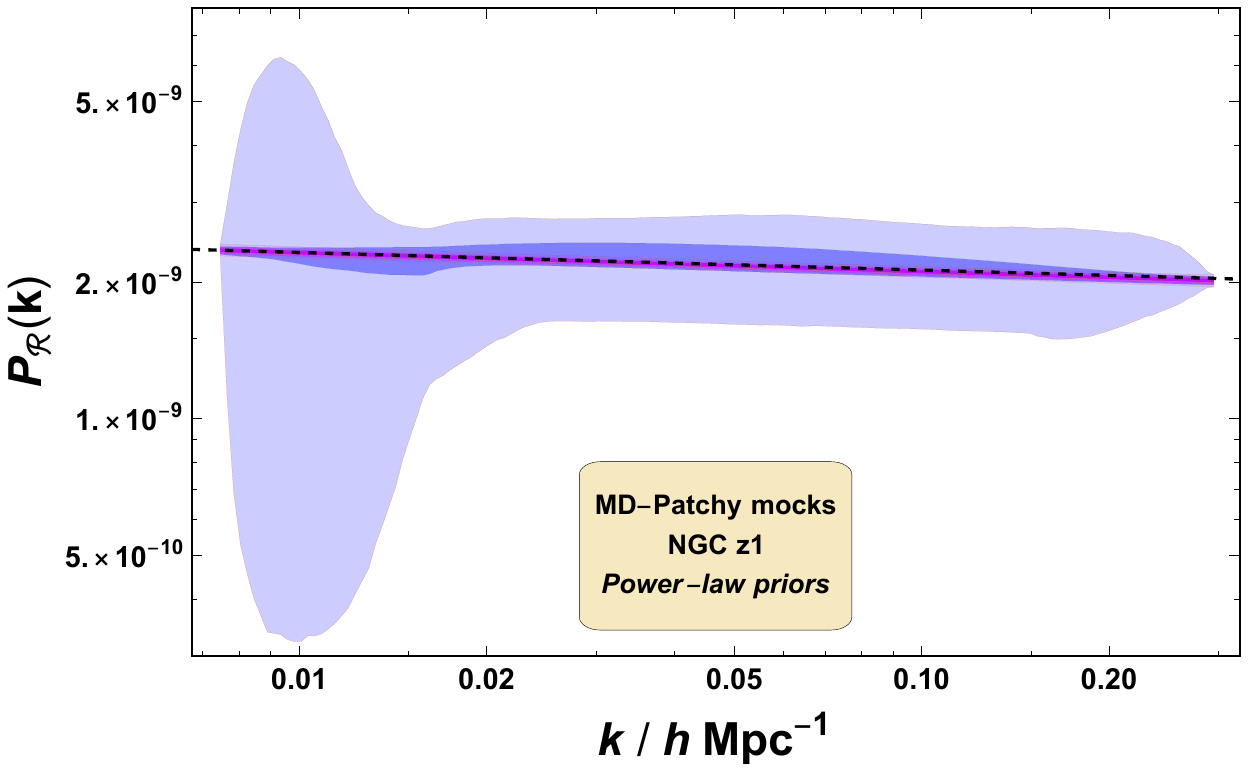}
\includegraphics[width=.33\textwidth]{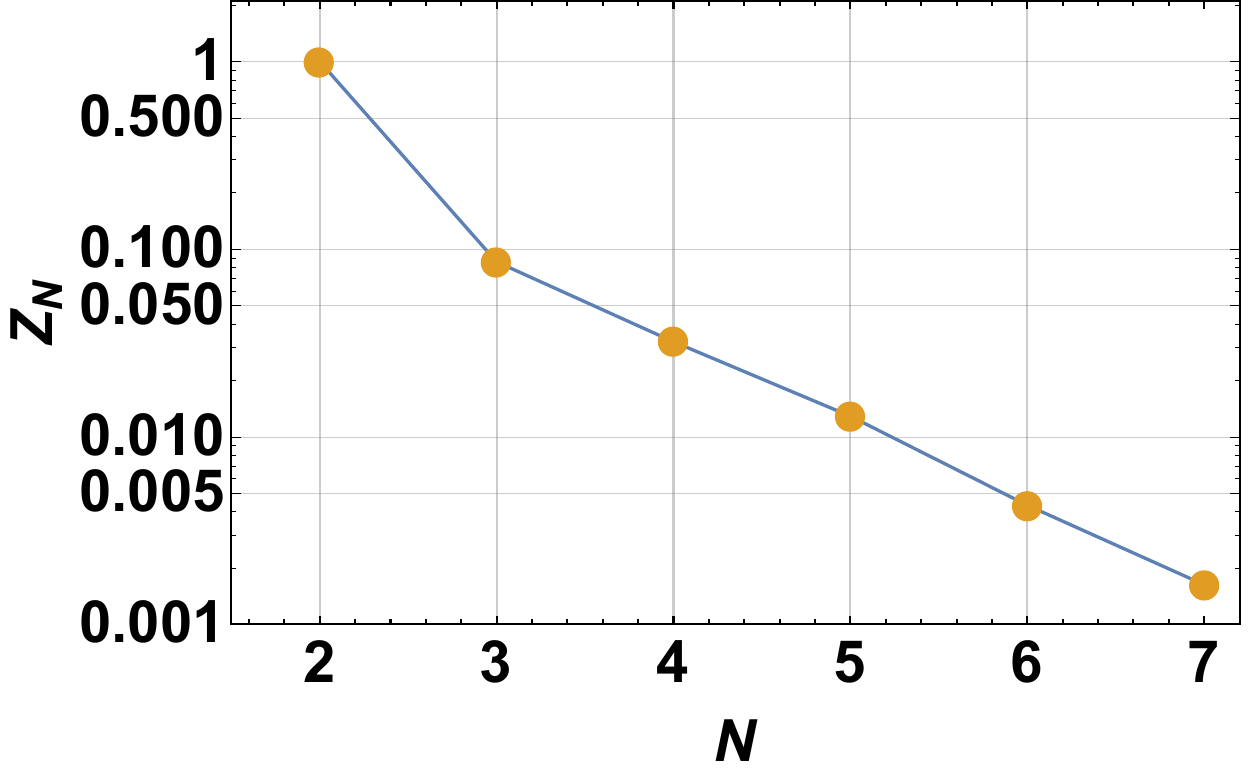}
\includegraphics[width=.50\textwidth]{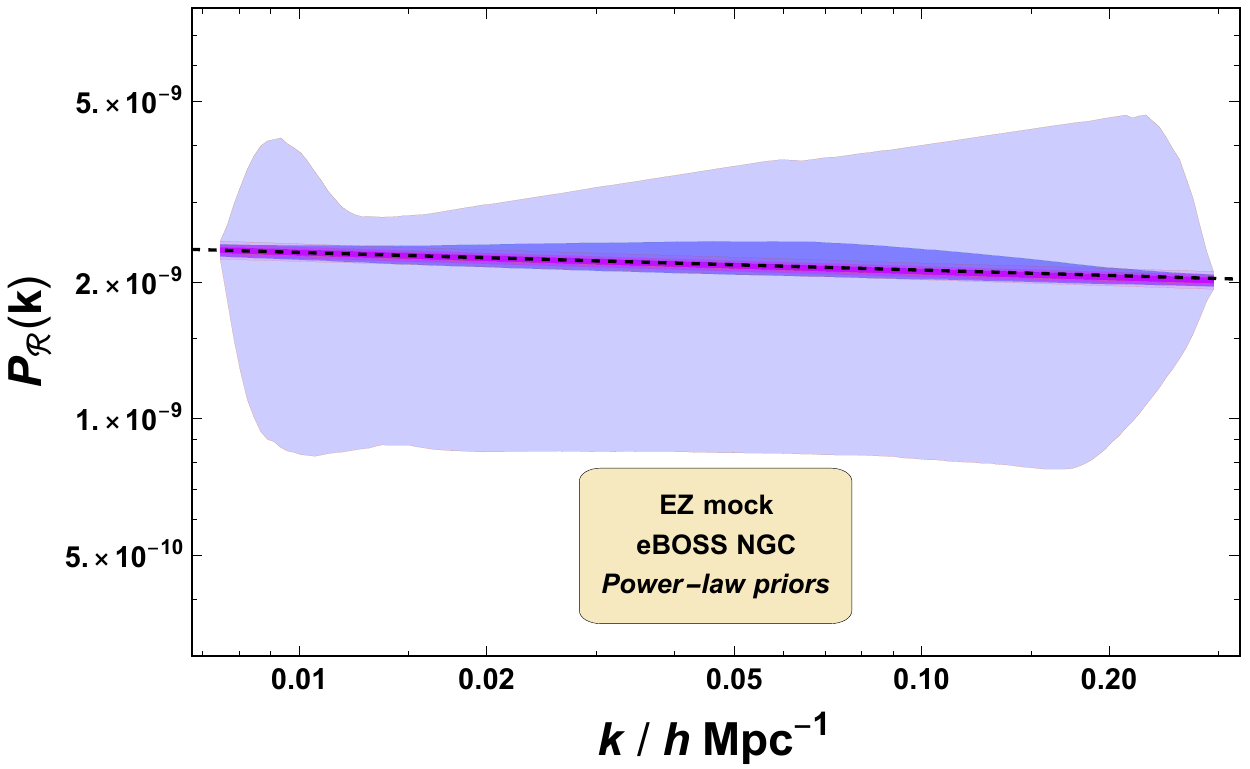}
\includegraphics[width=.33\textwidth]{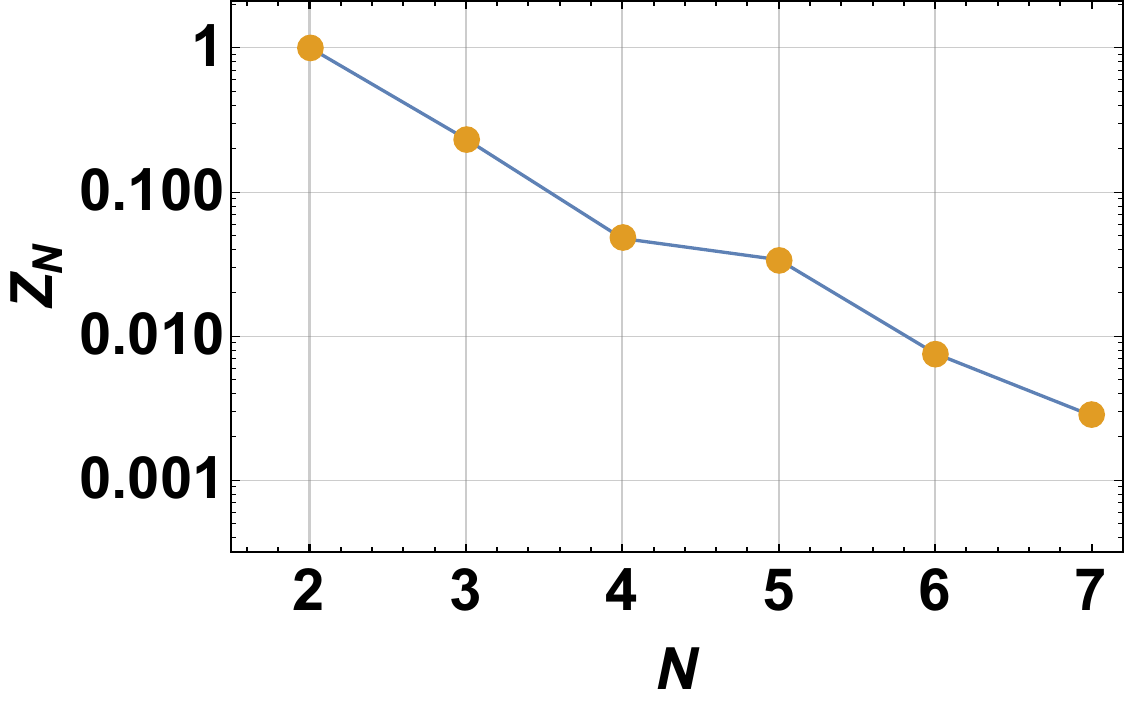}
\caption{$P_{\mathcal{R}}(k)$ reconstructions applied to the same BOSS and eBOSS mocks as in \cref{fig:MocksNGCz1}, but assuming the \textit{power-law} priors.}
\label{fig:MocksPowerLaw}
\end{figure}

We tested the PF sensitivity for this new prior set, obtaining better sensitivity to features than for the \textit{model priors}. The local oscillatory 5\% feature is recovered with decisive statistical significance when combining all the bins. For the 3\% LO and power-law templates we also obtain no feature evidence. Once the PF sensitivity is proved for this prior set, we performed the $P_{\mathcal{R}}(k)$ reconstructions to both BOSS and eBOSS mock catalogs, represented in \cref{fig:MocksPowerLaw}. The mocks' power-law is recovered, as with the \textit{model prior} set.

\begin{figure}[t]
\centering 
\includegraphics[width=.65\textwidth]{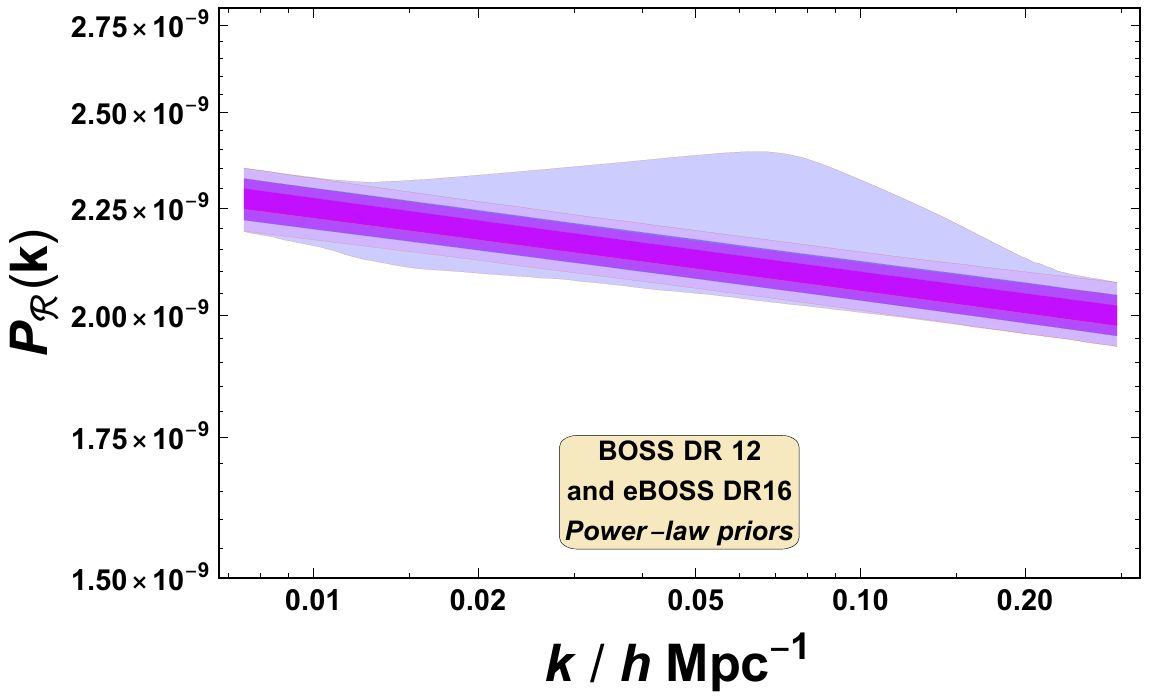}
\includegraphics[width=.35\textwidth]{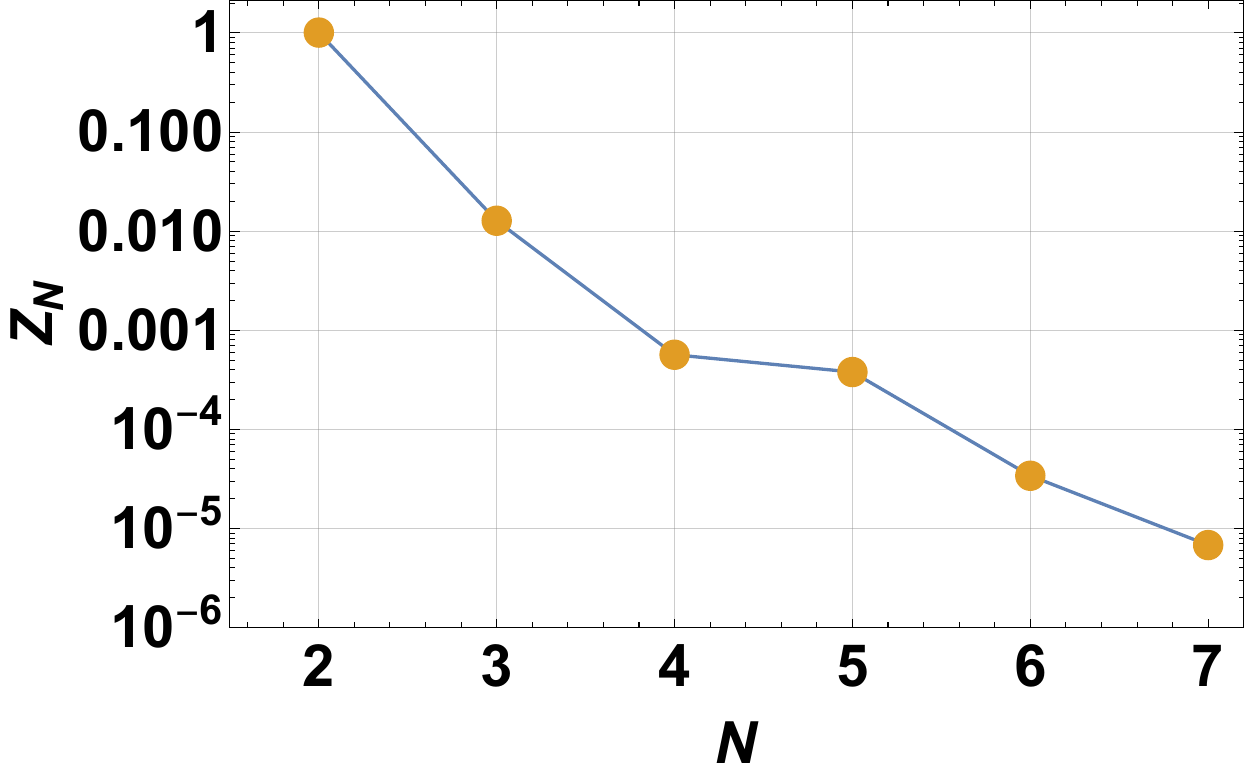}
\caption{Top: primordial power spectrum reconstruction for the combined BOSS DR 12 galaxies and eBOSS DR 16 QSO, assuming the \textit{power-law priors}. Bottom: evidence values for this scenario.}
\label{fig:CombinedPowerLawPriors}
\end{figure}

Since the validation with mocks yielded similar results as with the \textit{model priors}, we move on to reconstruct $P_{\mathcal{R}}(k)$ from data with the \textit{power-law priors}. Now, the obtained reconstructions incorporate Planck DR3 information through the knot amplitudes, while $\Theta_{\text{model}}$ is more agnostic. We show in \cref{fig:CombinedPowerLawPriors} the $P_{\mathcal{R}}(k)$ reconstruction for the combined $z$-bins and caps. As in the \textit{model priors} case, we obtain a clear preference for a power-law in all $z$-bins and caps, which is coherently reflected in the joint reconstruction. The Bayes factor for this combined scenario reaches a value of $79$, and as in the \textit{model priors} case, the $N$-marginalized and 2-knot reconstructions overlap, except at the $3\sigma$ level. A slight broadening appears around $k \approx 0.07 \text{ h} \text{ Mpc}^{-1}$, but it is not significant enough to suggest the presence of a PF. When comparing with the \textit{model priors} scenario, the \textit{power-law priors} are noticeably tighter, yielding an even more precise $P_{\mathcal{R}}(k)$ reconstruction. The inferred values of both $\{A_s, n_s\}$ are also compatible with Planck DR3.

Overall, the reconstructions obtained with both \textit{model priors} and \textit{power-law priors} show a clear and consistent preference for a primordial power-law spectrum. Despite the different assumptions on the knots and model parameters, the data favor a featureless $P_{\mathcal{R}}(k)$, with a power-law emerging as the statistically most favored description of $P_{\mathcal{R}}(k)$ across all bins and prior choices.

\bibliographystyle{JHEP.bst}
\bibliography{bibliography}

\end{document}